\documentclass[fleqn,usenatbib]{mnras}

\usepackage[T1]{fontenc}
\usepackage{ulem}
\usepackage{multicol}
\DeclareRobustCommand{\VAN}[3]{#2}
\let\VANthebibliography\thebibliography
\def\thebibliography{\DeclareRobustCommand{\VAN}[3]{##3}\VANthebibliography}

\usepackage{graphicx}	
\usepackage{amsmath}	

\newcommand{\msol}{M$_\odot$~}

\title[Constraining hyperonic dense matter property]{Constraining a relativistic mean field model using neutron star mass-radius measurements II: Hyperonic models}

\author[Chun Huang et al.]{
Chun Huang,$^{1}$\thanks{E-mail: chun.h@wustl.edu}
Laura Tolos,$^{2,3,4}$
Constan\c{c}a Provid\^{e}ncia,$^{5}$
and Anna Watts$^{6}$ 
\\
$^{1}$Physics Department and McDonnell Center for the Space Sciences, Washington University in St. Louis; MO, 63130, USA \\
$^{2}$Institute of Space Sciences (ICE, CSIC), Campus UAB, Carrer de Can Magrans, 08193, Barcelona, Spain \\
$^{3}$Institut d'Estudis Espacials de Catalunya (IEEC), 08860 Castelldefels (Barcelona), Spain \\
${^4}$Frankfurt Institute for Advanced Studies, Ruth-Moufang-Str. 1, 60438, Frankfurt am Main, Germany \\
$^{5}$CFisUC, Department of Physics, University of Coimbra, 3004-516 Coimbra, Portugal \\
$^{6}$Anton Pannekoek Institute for Astronomy, University of Amsterdam, Science Park 904, 1090 GE Amsterdam, the Netherlands\\
}

\date{Accepted XXX. Received YYY; in original form ZZZ}

\pubyear{2024}

\begin{document}
\label{firstpage}
\pagerange{\pageref{firstpage}--\pageref{lastpage}}
\maketitle

\begin{abstract}

We investigate whether measurements of the neutron star mass and radius or the tidal deformability can provide information about the presence of hyperons inside a neutron star. This is achieved by considering two inference models, with and without hyperons, based on a field-theoretical approach. While current observations do not distinguish between the two scenarios, we have shown that data simulating expected observations from future large area X-ray timing telescopes could provide some information through Bayes factors. Inference using simulated data generated from an EOS containing hyperons decisively favours the hyperonic model over the nucleonic model. However, a 2\% uncertainty in the mass and radius determination may not be sufficient to constrain the parameters of the model when only six neutron star mass-radius measurements are considered.
\end{abstract}

\begin{keywords}
dense matter - equation of state - stars: neutron - X-rays: general
\end{keywords}


\section{Introduction}
\label{intro}

If stable states of strange matter exist anywhere in the Universe, the most likely location is in the cores of neutron stars, where densities reach several times the nuclear saturation density \citep{Chatterjee:2015pua,Tolos2020,Burgio2021}. Strange matter could take various forms (deconfined quarks, mesons), but one possibility is that it is baryonic, in the form of {\it hyperons} \citep{Ambartsumyan60}.   

The presence of hyperons would soften the dense matter Equation of State (EOS), affecting the overall structure of the star and the relationship between neutron star mass and radius/tidal deformability. The presence of hyperons can therefore in principle be explored using new observational techniques. Constraints on mass and tidal deformability can be derived from the properties of gravitational waves (GW) emitted during the final stages of neutron star binary inspiral \citep{2018PhRvL.121p1101A,LIGOScientific:2020aai}.  
And pulse profile modeling, using X-ray data from NICER \citep[the Neutron Star Interior Composition Explorer,][]{Gendreau2016}, allows the inference of neutron star mass and radius \citep{Riley2019,Miller2019,Riley2021,Miller2021,Salmi22,Salmi23,Vinciguerra24,Salmi24,Dittmann24,Choudhury24,Salmi24b}. Applying the latter technique to faint rotation-powered millisecond pulsars and accreting neutron stars is a key science driver for planned and proposed future X-ray telescopes \citep{2013sf2a.conf..447B,Watts_2016,extp_watts, strobex2024}.  

These new observational constraints are already being applied to inform our understanding of dense matter \citep{Miller2021,Raaijmakers_2021, Annala23,Takatsy23,Pang24,Koehn24}.  Many of these studies use EOS meta-models (which may be parameterized or non-parameterized) that attempt to span all reasonable mass-radius parameter space, rather than being driven by microphysics \citep{Kurkela:2014vha,Annala18,Tews2018,Annala2020,Landry:2020vaw,Essick2021,Legred21,Altiparmak:2022bke,Gorda:2022jvk,Rutherford24}.  This means that that they provide no direct insight into composition. In this study we will explore the constraints that current NICER and GW results pose for hyperonic models, using a microphysical model. We will also explore the prospects for improved constraints from the next generation of X-ray telescopes, which are expected to generate tighter mass-radius inferences via pulse profile modeling.  We use a microscopic model
based on a relativistic field theoretical approach for the EOS, which allows us to explore composition, and our sensitivity to the possible presence of hyperons, directly.  This study builds on earlier work using a purely nucleonic model \citep{Huang24}.

\section{Equation of State Models}
\label{EOS}

\begin{table*}
\begin{tabular}{ccccccccccc}
\hline \hline \text { Model } & $m_{\sigma}$ & $m_{\omega}$ & $m_{\rho}$ & $g_{\sigma N}^{2}$ & $g_{\omega N}^{2}$ & $g_{\rho N}^{2}$ & $\kappa$ & $\lambda$ & $\zeta$ & $\Lambda_{\omega}$ \\
& $(\mathrm{MeV}) $& $(\mathrm{MeV})$ & $(\mathrm{MeV})$ & & & & & & & \\
\hline 
\text { FSU2R } & 497.479 & 782.500 & 763.000 & 107.5751 & 182.3949 & 206.4260 & 3.0911 & -0.001680 & 0.024 & 0.045 \\
\text { FSU2H } & 497.479 & 782.500 & 763.000 & 102.7200 & 169.5315 & 197.2692 & 4.0014 & -0.013298 & 0.008 & 0.045 \\
\text { TM1-2$\omega\rho$ } & 511.198 & 783.000 & 770.000 & 99.9661 & 156.3384 & 127.7469 & 3.5235 & -0.004739 & 0.012 & 0.030 \\
\hline \hline
\end{tabular}
\caption{Parameters of the three EOS considered in the present work.}
\label{tab-par}
\end{table*}

We base our analysis on an EOS coming from a relativistic mean field (RMF) description of hadronic matter. In particular, the calibrated parameter set of the FSU2R EOS \citep{Tolos_2016,Tolos_2017} is chosen as the central value of our prior distributions for the nucleonic parameters, as done in our previous work \citep{Huang24}.  This RMF scheme for nucleons was optimised to describe 2\msol stars and to satisfy the condition that the radius is less than 13 km, while reproducing the properties of nuclear matter and finite nuclei as well as certain  restrictions on high-density matter deduced from heavy-ion collisions. In the present paper, we also take into account the effect of hyperons, thus taking a similar approach as the FSU2H parameterisation of \cite{Tolos_2016,Tolos_2017}. The hyperonic parameters are then calculated by fitting the experimental data available for hypernuclei, in particular, the possible values of the optical potential of hyperons  extracted from these data. The parameters of the FSU2R and FSU2H are shown in the first two rows of Table~\ref{tab-par}. Note that there are many EOSs in the literature that are based on the same RMF framework, i.e. the same underlying Lagrangian density. Although our prior was defined with respect to a particular EOS, we could have started from any other EOS.

In the next subsection, we focus on how to introduce the hyperon sector on top of the nucleonic EOS model. We will not describe the nucleonic part of RMF theory in detail here, as we have done that in our previous work, see \citet{Huang24}.  	

\subsection{Hyperonic Equation of State}
\label{sub:hyperons}

In order to discuss the role of hyperons inside neutron stars within the RMF framework, it is necessary to determine the hyperon-meson couplings. So we start by recalling that the potential felt by hyperon $i$ on the $j$ particle matter at density $n_j$ is given by
\begin{equation}
U_{i}^{(j)}\left(n_{j}\right)=-g_{\sigma i} \bar{\sigma}^{(j)}+g_{\omega i} \bar{\omega}^{(j)}+g_{\rho i} I_{3 i} \bar{\rho}^{(j)}+g_{\phi i} \bar{\phi}^{(j)}
\label{Ypot}
\end{equation}
where $g_{\sigma i}$,  $g_{\omega i}$, $g_{\rho i}$ and $g_{\phi i}$ are the couplings of the $\sigma$, $\omega$, $\rho$ and $\phi$ fields to hyperons, and $\bar{\sigma}^{(j)}, \bar{\omega}^{(j)}, \bar{\rho}^{(j)}$ and $\bar{\phi}^{(j)}$ are the meson mean-fields in the $j$-particle matter.  
%
			
Flavor SU(3) symmetry, the vector dominance model and ideal mixing for the physical $\omega$ and $\phi$ mesons allows one to relate the couplings between the hyperons and the $\omega$ and $\phi$ mesons to the nucleon coupling $g_{\omega N}$ as in \cite{Schaffner:1995th,Weissenborn:2011ut,Providencia:2012rx,Miyatsu:2013hea,Banik:2014qja,Tolos_2016,Tolos_2017}, that is, 
\begin{eqnarray}
g_{\omega \Lambda}:g_{\omega \Sigma}:g_{\omega \Xi}:g_{\omega N}&=&\frac{2}{3}:\frac{2}{3}:\frac{1}{3}:1 \nonumber \\
g_{\phi \Lambda}: g_{\phi\Sigma}: g_{\phi\Xi}:g_{\omega N}&=& -\frac{\sqrt{2}}{3}: -\frac{\sqrt{2}}{3}:  -\frac{2\sqrt{2}}{3}: 1 , \ \ \ \ \ \ \ 
\label{eq:couplings}
\end{eqnarray}
where we note that $g_{\phi N}=0$. The coupling of the  $\Lambda$ hyperon to the $\phi$ meson is however reduced by 20\% so as to obtain a $\Lambda\Lambda$ bond energy in $\Lambda$ matter at a density $n_\Lambda \simeq n_0/5$ of $\Delta B_{\Lambda\Lambda} (n_0/5) = 0.67$ MeV, thus reproducing the value extracted from the 
$^6_{\Lambda\Lambda} {\rm He}$ double $\Lambda$ hypernucleus \citep{Takahashi:2001nm, E373KEK-PS:2013dfg}. As for the couplings of hyperons to $\rho$ we relate them to $\rho N$ according to
\begin{eqnarray}
g_{\rho \Lambda}:g_{\rho \Sigma}:g_{\rho \Xi}:g_{\rho N}&=&0:1:1:1 \nonumber .
\label{eq:couplings2}
\end{eqnarray}
We should indicate that the isospin operator $I_{3 i} $ appearing in the definition of the potentials in Eq.~(\ref{Ypot}) implements the relative factor of 2 missing in the 1:1 relation between  $g_{\rho \Sigma}$ and $g_{\rho N}$ displayed in Eq.~(\ref{eq:couplings2}) due to isospin, so that the effective coupling of the $\rho$ meson to the $\Sigma$ hyperon ($I_{3}=-1,0,+1$)  is twice that to the nucleon ($I_{3}=-1/2,+1/2$), as required by the isospin symmetry.

Then, the hyperon-$\sigma$ coupling can be determined from Eq.~(\ref{Ypot}) by reproducing the hyperon optical potential in symmetric nuclear matter at saturation density $n_0$, as derived from hypernuclear data. For the $\Lambda$ potential, the Woods-Saxon type potential of $U_\Lambda^{(N)}(n_0) \sim -28$ MeV reproduces the bulk of $\Lambda$ hypernuclei binding energies \citep{Millener:1988hp}. With regards to the $\Sigma$ hyperon, a moderate repulsive potential could be extracted from $(\pi^-, K^+)$ reactions off nuclei \citep{Noumi:2001tx} done in \citep{Harada:2006yj,Kohno:2006iq}.  Fits to $\Sigma^-$ atomic data \citep{Friedman:2007zza} also indicate a transition from an attractive $\Sigma$-nucleus potential at the surface to a repulsive one inside the nucleus, but  the repulsion is not well determined. As for $\Xi$, the potential in symmetric nuclear matter is also quite uncertain. Whereas emulsion data shows  attractive values of  $U_\Xi^{(N)}(n_0) = - 24 \pm 4$ MeV \citep{Dover:1982ng}, 
analyses of the $(K ^-, K^+)$ reaction on a $^{12}$C target indicate a milder attraction \citep{E224:1998uzu,AGSE885:1999erv}. More recently, the value of $U_{\Xi}=-21.9\pm0.7$ MeV has been determined in \cite{Friedman:2021rhu}.  This value was updated to -13.8 MeV by the same authors \citep{Friedman:2022huy} using recent data obtained by experience E07 at J-PARC \citep{Yoshimoto:2021ljs}.  Therefore, we take into account these experimental uncertainties so as to vary the hyperon potentials in symmetric nuclear matter  as follows:
 \begin{eqnarray}
U_{\Lambda}^{(N)}(n_0)&=& -25~{\rm to}~ -30 ~{\rm MeV}  \nonumber \\
U_{\Sigma}^{(N)}(n_0)&=&10~{\rm to}~ 40~{\rm MeV}  \nonumber \\
U_{\Xi}^{(N)}(n_0)&=&-10~{\rm to}~-25~{\rm MeV}  \ .
\label{eq:pots}
\end{eqnarray}
This range of values for the hyperon potentials at $n_0$ gives rise to the following range for the hyperon-$\sigma$ couplings:
\begin{eqnarray}
g_{\sigma \Lambda}/g_{\sigma N}&=& 0.60475 - 0.61783  \nonumber \\
g_{\sigma  \Sigma}/g_{\sigma N}&=& 0.43470 - 0.51319  \nonumber \\
g_{\sigma  \Xi}/g_{\sigma N} &=& 0.29583 - 0.33507   \ ,
 \label{eq:sigma} 
\end{eqnarray}
where the lower values correspond to the most repulsive situation  and the upper ones to the most attractive one. 

Note that this way of proceeding is similar to the one discussed for the FSU2H parameterization in ~\cite{Tolos_2016, Tolos_2017}. The difference arises regarding the limiting allowed values for the hyperon potentials, as they are not well constrained for the case of $\Sigma$ and $\Xi$.

To finalize this section, let us point out that the  SU(3) symmetry and  vector dominance have constrained the vector meson sector, with the exception of the coupling of the $\phi$ to $\Lambda$. In addition, the SU(3) symmetry was not assumed for the scalar-baryon couplings, since they are determined by hypernuclei data. While for the $\Lambda$ and $\Xi$ hyperons an approximate SU(3) symmetry is obtained, this is not the case with the $\Sigma$-hyperon. 
In several studies  the SU(3) symmetry was not considered in general, and stiffer EOSs were obtained, including \cite{Bednarek:2011gd,Weissenborn:2011ut,Colucci:2013pya,Lopes:2013cpa,Oertel:2014qza}. In fact, the breaking of the flavor SU(3) symmetry to produce stiffer EoSs has been advocated as a possible way to solve the hyperon puzzle in neutron stars.
			
\subsection{Choice of priors for model parameters}
\label{prior}
\begin{table}
\centering
\setlength{\tabcolsep}{11mm}{\begin{tabular}{l c} 
\hline\hline
\text {EOS parameter} & {Prior} \\
\hline
$\kappa$ (MeV)&$N(2.525, 1.525^2)$\\ $\lambda_{0}$ & $N(0.0045,0.0205^2)$\\
$\zeta$ & $\mathcal{U}(0,0.04)$\\
$\Lambda_{\omega}$&$\mathcal{U}(0,0.045)$\\ 
$g_{\sigma  }^{2}$&$N(107.5, 7.5^{2})$\\
$g_{\omega }^{2}$& $\mathcal{U}(150, 210)$\\ $g_{\rho }^{2}$&$ \mathcal{U}(75,210)$\\
[1ex] 
\hline\hline
$g_{\sigma \Lambda} / g_{\sigma N}$& $N(0.61129, 0.01308^{2})$\\
                 $g_{\sigma \Sigma} / g_{\sigma N}$& $N(0.47395, 0.07849^{2})$\\
                 $g_{\sigma \Xi} / g_{\sigma N}$&
                 $N(0.31545, 0.03924^{2})$\\ 
                 $g_{\phi \Lambda} / g_{\omega N}$&
                 $N(0.87407, 0.10447^{2})$\\
            [1ex] 
\hline\hline    
\end{tabular}}
\caption{The prior distributions assumed for the EOS parameters, where $N$ is a Gaussian distribution and $\mathcal{U}$ a Uniform (Flat) distribution.}
\label{table:1}
\end{table}

 The choice of priors for the present analysis is given in Table~\ref{table:1}. For the nucleonic parameters, $g_{\sigma}$, $g_{\omega}$, $g_{\rho}$,  which define, respectively, the strength of the coupling of the $\sigma$, $\omega$ and $\rho$ mesons to the nucleons, and  $\kappa$, $\lambda_0$, $\zeta$ and $\lambda_{\omega}$ which determine the strength of the meson self-interacting and mixed terms, we use the same central values and priors as in our previous work of \cite{Huang24} (see Eqs. (1) and (2)for the Lagrangian density in the former reference).
For the newly introduced hyperonic degrees of freedom, the hyperon potentials at saturation density determine the range of the $\sigma$-hyperon couplings, while the $\Lambda$ potential at saturation  and the $\Lambda \Lambda$ bond at lower densities fix the value of the $\phi$-$\Lambda$ coupling, as discussed in Sec.~\ref{sub:hyperons}. 

It is critical for the inference method to have a prior distribution broad enough to encompass all possibilities. We therefore assume the $g_{\sigma \Lambda} / g_{\sigma N}$,  $g_{\sigma \Sigma} / g_{\sigma N}$ and $g_{\sigma \Xi} / g_{\sigma N}$ priors to be Gaussian distributions whose 68\% confidence intervals are within the limiting cases determined in Eq.~(\ref{eq:sigma}). For the  $g_{\phi \Lambda} / g_{\omega N}$ the 68\% credible interval lies within 0.821 < $g_{\phi \Lambda} / g_{\omega N}$ < 0.926, which results from modifying the coupling of the $\Lambda$ hyperon to the $\phi$ meson for a given $g_{\sigma \Lambda} / g_{\sigma N}$ so as to obtain the $\Lambda\Lambda$ bond energy.

Taking into account all the nucleonic and hyperonic parameters, the nucleonic model space will be a seven dimensional parameter space, while the consideration of the hyperon sector introduces another four dimensions, giving rise to an eleven dimensional parameter space.

\subsection{Nuclear matter saturation properties}

Following the method described in \citet{Chen_2014}, the EOS parameters can be related to the saturation properties of nuclear matter. Without repeating all the derivations here, the quantity of particular interest to us is the incompressibility of symmetric nuclear matter $K$, since it has a much wider range of possible values than the experiments indicate. 
To control it, a minimal guidance from nuclear physics is included as in \cite{Huang24}, imposing the probability function $p(K) = -0.5\times|250 - K|^{10}/150^{10}$ to be the prior filter that shapes the prior before we implement it in the inference. For the other nuclear quantities, we fix $E/A$ to be a Gaussian distribution centred on -16 MeV, with $\sigma = 0.02$, $N(-16, 0.02^{2})$, and restrict $m^{*}$ to $\mathcal{U}(0.55,0.64)$ and $n_0$ to $\mathcal{U}(0.15,0.17)$, since they are sufficiently well constrained by nuclear experiments.

\subsection{Mass-radius priors}
\begin{figure}
\includegraphics[width=\linewidth]{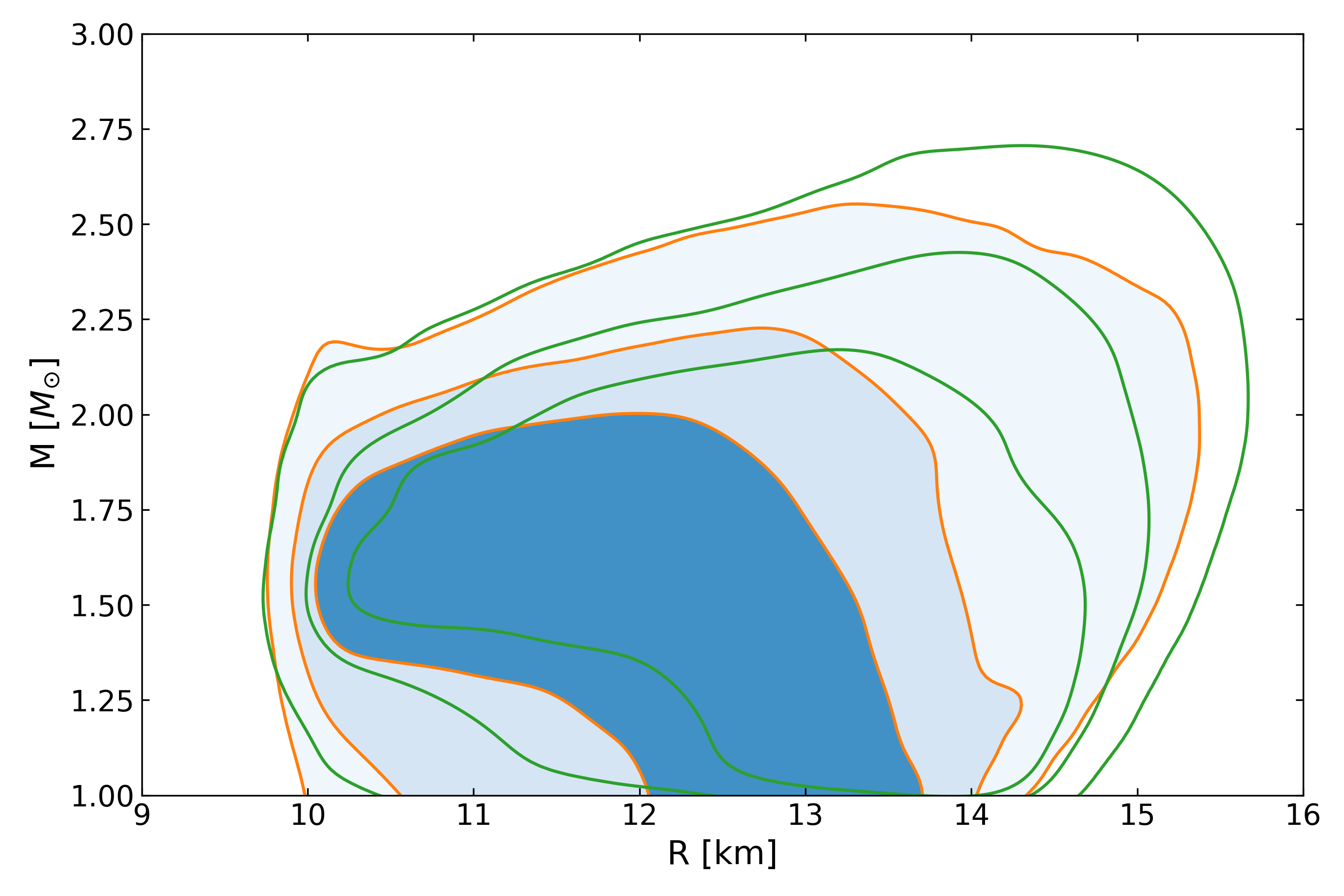}
\caption{The M-R posterior after applying the nuclear constraint (orange/blue), compared to the M-R prior resulting from the initial EOS model and priors (green). The contour levels, from the innermost to the outermost, correspond to the 68\%, 84\% and 100\% credible regions, where 100\% is the point beyond which there are no samples (for both the orange/blue and green contours). }
\label{M-R_nu} 
\end{figure}
The EOS can be mapped into the neutron star Mass-Radius (M-R) domain by solving the Tolman-Oppenheimer-Volkoff (TOV) equations
\citep{Tolman:1939jz,Oppenheimer:1939ne}. The TOV equations for a  static and spherically-symmetric star are given by
			\begin{equation}
			\begin{aligned}
			\frac{d P}{d r} &=-\frac{G}{r^{2}}(\varepsilon+P)\left(m+4 \pi r^{3} P\right)\left(1-\frac{2 G m}{r}\right)^{-1}, \\
			\frac{d m}{d r} &=4 \pi r^{2} \varepsilon ,
			\end{aligned}
			\end{equation}
			where $P$, $\varepsilon$ and  $m$ are the pressure, the energy density and  the star mass, respectively, for a given radius $r$, the radial coordinate in spherical coordinates, and $G$ is the gravitational constant. 

In order to solve the TOV equations, the EOS for the whole neutron star is needed. The EOS described above by the RMF method only describes the core. For the crust, the BPS outer crust EOS \citep{Baym:1971pw} was applied for $\varepsilon < \varepsilon_{\text {outer }} = 4.30 \times 10^{11} \mathrm{~g} / \mathrm{cm}^{3}$. At the interface between the core and the outer crust, $\varepsilon_{\text {outer }} < \varepsilon<\varepsilon_{c} =2.14 \times 10^{14} \mathrm{~g} / \mathrm{cm}^{3}$ region, we have implemented the same treatment as in \cite{Huang24}, that is, a polytrope fitting that includes four additional points from the unified inner crust EOS obtained in \cite{Providencia:2018ywl} to improve the fit. A detailed discussion of this method can be found in our previous work \citep{Huang24}.
			
We show in Figure \ref{M-R_nu} the hyperonic NS mass-radius prior, defined by the green contours, resulting from the eleven dimensional parameter prior space (nucleon + hyperon) given in Table \ref{table:1}. 
Our choice of priors does not include samples larger than 16 km, but does allow radii smaller than 10 km (the minimum allowed radius is 9.8 km). As a comparison, the full prior for the nucleonic model used in \citet{Huang24} (see Figure 2 of that paper)\footnote{The nucleonic EOS prior we use here is still consistent with \cite{Huang24} (see Figure 1 of that paper).} allowed radii larger than 16 km, but the radius was limited by applying nuclear constraints (the minimum allowed radius for the constrained nucleonic model prior was 10.1 km, and the maximum 15.5 km). After filtering the initial hyperonic model prior with the nuclear constraints defined above for $K$, $E/A$, $m^*$ and $n_0$, the radius still extends to below 10 km (orange/blue contours in Figure~\ref{M-R_nu}), which matches expectations, as the introduction of hyperons will tend to soften the EOS.    
            
The maximum mass is smaller once nuclear constraints are imposed. The presence of hyperons tends to soften the EOS, producing generally lower mass stars for the same central density. This effect is clearly seen by comparing the maximum mass star of the constrained hyperonic prior ($\sim 2.5$ \msol) with that of the constrained nucleonic prior ($\sim 3.0$ \msol). 
          
		\section{Inference framework}
  \label{inference}
  \subsection{Inference method}
The inference conducted here relies on the framework developed by the authors, namely the \textit{CompactObject} \citep{EoS_inference} package. This is an open source full-scope package  designed to implement Bayesian constraints on the neutron star EOS.

Two types of inference methods will be explored: i) current observations of neutron stars (maximum masses derived from radio pulsar timing, GW measurements of tidal deformability, and M-R measurements from NICER), and ii) future mass-radius measurements simulated for future large-area X-ray telescopes. Our goal with these inferences is to test whether current observations can place any constraints on the parameters of the hyperonic model, and to investigate to what extent we can probe the existence of hyperons inside neutron stars using future X-ray telescopes.
        
The Bayesian inference methodology described here follows the same framework developed by \citet{Greif_2019} and \citet{Raaijmakers_2019,Raaijmakers_2020,Raaijmakers_2021}. Bayes' theorem gives the nuisance-marginalized likelihood function as
\begin{equation}
p(\boldsymbol{\theta}, \varepsilon \mid \boldsymbol{d}, \mathcal{M}) \propto p(\boldsymbol{\theta} \mid \mathcal{M}) p(\varepsilon \mid \boldsymbol{\theta}, \mathcal{M}) p(\boldsymbol{d} \mid \boldsymbol{\theta}, \mathcal{M}) ,
\label{like}
\end{equation}
where $p(\boldsymbol{\theta},\varepsilon\mid\boldsymbol{d},\mathcal{M})$ is the posterior distribution of  $\boldsymbol{\theta}$ and central energy densities $\varepsilon$. 
For the nucleonic model, $\boldsymbol{\theta}$ is a 7-dimensional vector, whereas for the present hyperonic model  this vector spans eleven dimensions (7-d nucleon + 4-d hyperon d.o.f). The variable $\mathcal{M}$ indicates the model we use, and $\boldsymbol{d}$ stands for the dataset. 
\subsection{Observational data sets}            

For current constraints, we can consider the following types of observations: measurements of high masses for radio pulsars; joint M-R measurements inferred from NICER data, and mass-tidal deformability measurements from GW events (see Section \ref{results_now} for specifics of the observational data sets used). Since these measurements are independent, the likelihood function can be written as
\begin{equation}
\begin{aligned}
&p(\boldsymbol{\theta}, \varepsilon \mid \boldsymbol{d}, \mathcal{M}) \propto p(\boldsymbol{\theta} \mid \mathcal{M}) p(\varepsilon \mid \boldsymbol{\theta}, \mathcal{M})\\
&\times \prod_{i} p\left(\Lambda_{1, i}, \Lambda_{2, i}, M_{1, i}, M_{2, i} \mid d_{\mathrm{GW}, \mathrm{i}}\left(, \boldsymbol{d}_{\mathrm{EM}, \mathrm{i}}\right)\right)\\
&\times \prod_{j} p\left(M_{j}, R_{j} \mid d_{\mathrm{NICER}, \mathrm{j}}\right)\\
&\times \prod_{k} p\left(M_{k} \mid \boldsymbol{d}_{\mathrm{radio, \textrm {k }}}\right).
\end{aligned}
\end{equation}
We further modify the GW posterior distributions to include the two tidal deformabilities, the chirp mass and the mass ratio $q$, reweighting them simultaneously. The posterior formula then becomes
\begin{equation}
\begin{aligned}
&p(\boldsymbol{\theta}, \varepsilon \mid \boldsymbol{d}, \mathcal{M}) \propto p(\boldsymbol{\theta} \mid \mathcal{M}) p(\varepsilon \mid \boldsymbol{\theta}, \mathcal{M}) \\
&\quad \times \prod_{i} p\left(\Lambda_{1, i}, \Lambda_{2, i}, q_{i} \mid \mathcal{M}_{c}, \boldsymbol{d}_{\mathrm{GW}, \mathrm{i}}\left(, \boldsymbol{d}_{\mathrm{EM}, \mathrm{i}}\right)\right) \\
&\quad \times \prod_{j} p\left(M_{j}, R_{j} \mid \boldsymbol{d}_{\mathrm{NICER}, \mathrm{j}}\right) \\
&\quad \times \prod_{L} p\left(M_{k} \mid \boldsymbol{d}_{\text {radio } \mathrm{k}}\right) .
\end{aligned}
\end{equation}
Here $\Lambda_{2,i}=\Lambda_{2,i}(\boldsymbol{\theta} ; q_{i})$ is the tidal deformability. We follow the same convention as in \cite{2018PhRvL.121p1101A} and assume $M_{1}>M_{2}$, since the GW probability is symmetric when changing $M_1$ to $M_2$ or vice versa.

\subsection{Simulated future observations}  
For future constraints, the data set $\boldsymbol{d}$ defined here consists of simulated M-R measurements with the precision that we expect from the next generation of X-ray telescopes (M-R credible regions should scale with exposure time and telescope effective area, see \citealt{Lo2013,Psaltis2014}). As in \citet{Huang24}, we do not include future GW and radio constraints in our future constraint scenario simulations, since we are interested in what can be achieved from mass-radius measurements alone.  The likelihood function in Eq.~(\ref{like}) is thus given by

\begin{equation}
\begin{aligned}
&p(\boldsymbol{\theta}, \varepsilon \mid \boldsymbol{d}, \mathcal{M}) \propto p(\boldsymbol{\theta} \mid \mathcal{M}) p(\varepsilon \mid \boldsymbol{\theta}, \mathcal{M}) \\
&\times \prod_{j} p\left(M_{j}, R_{j} \mid d_{\mathrm{STROBEX/eXTP}, j}\right) .
\end{aligned}
\end{equation}

First we need to generate simulated M-R measurements - injected data - assuming a "true" EOS of the neutron star such that the simulated measurements are based on an assumed M-R curve. We will consider a M-R curve from an EOS that contains only nucleonic degrees of freedom (with the associated simulated M-R measurements referred to as injected nucleonic data, or `inject-nucl' ) and another from an EOS  with nucleonic and hyperonic degrees of freedom (with the associated simulated M-R measurements referred to as injected hyperonic data or `inject-hyp').  When doing inference on this simulated data, we use two different models based on a microscopic RMF description of hadronic matter, one including only nucleonic degrees of freedom, which we identify as the nucleonic model, and a second one including nucleonic and hyperonic degrees of freedom, which we identify as the hyperonic model. 
These two models will be used to model the injected data to test whether our inference could distinguish or obtain evidence about the composition of neutron stars (with/without hyperons). We will have to perform four different Bayesian analyses: inject-nucl data with the nucleonic and hyperonic models, and inject-hyp data with the nucleonic and hyperonic models.

Two different injected data sets are generated, with simulated M-R data constructed from the RMF TM1-2$\omega\rho$n EOS and TM1-2$\omega\rho$nH EOS. The parameters of TM1-2$\omega\rho$ are given in the last row of Table~\ref{tab-par}. The TM1-2$\omega\rho$n EOS  defines the `inject-nucl' data and includes only nucleons (it was also used in our last paper, \citealt{Huang24}). The TM1-2$\omega\rho$nH EOS defines the `inject-hyp' data and includes hyperons and nucleons. 

In this study, our focus will be exclusively on the presentation of the `Future-X' scenario outlined in \cite{Huang24}, i.e. considering 2\% uncertainty M–R observational data. We proceed in this way in order to facilitate a comprehensive comparison between the two models under consideration, the nucleonic and the hyperonic models, for the two injected data sets, `inject-nucl' and `inject-hyp'.

\begin{figure}
	\centering
	\includegraphics[width = \columnwidth]{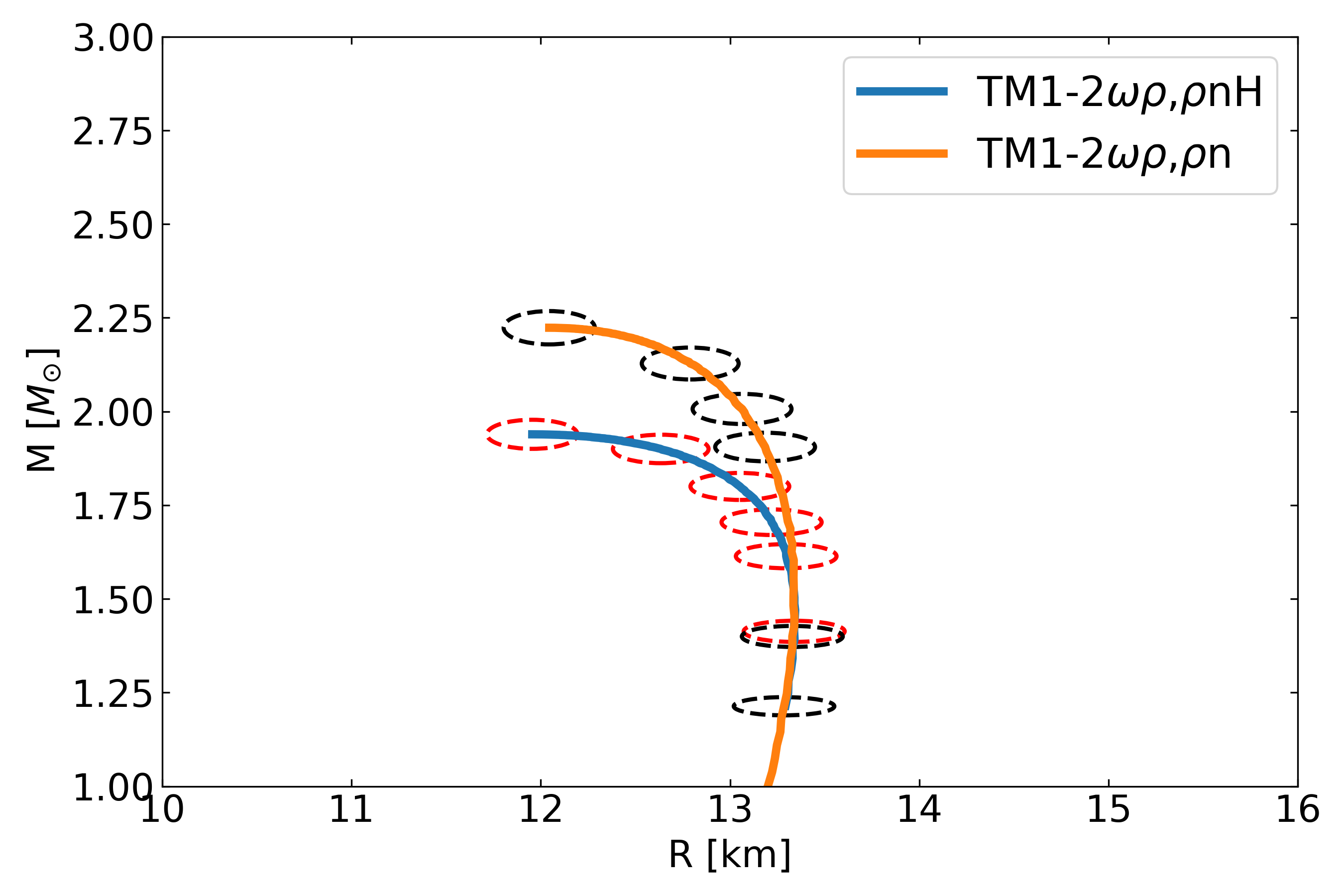}
	\caption{ The M-R curves determined from the TM1-2$\omega\rho$n EOS (without hyperons, orange line) and  the TM1-2$\omega\rho$nH EOS (with hyperons, blue line), which are the underlying EOS used to generate the simulated mass radius measurements for the `Future-X' scenarios. The red dashed curves are the six simulated M-R inject-hyp data from the TM1-2$\omega\rho$nH EOS, centered at [1.40, 1.60, 1.70, 1.80, 1.90, 1.94] \msol. The black dashed curves are the six simulated M-R inject-nucl data from the TM1-2$\omega\rho$n EOS, centered at [1.20, 1.40, 1.90, 2.00, 2.10, 2.20] \msol corresponding to the `Future-X' scenario with six 2\% uncertainty M-R observations. }
	\label{fig:ground_truth}
\end{figure}

Since the hyperonic injected data spans a smaller range of masses than in the nucleonic case, in the `Future-X' hyperon scenario described by the `inject-hyp' data we assume that we have six M-R measurements of neutron stars with 2\% uncertainty, whose masses are centered at [1.40, 1.60, 1.70, 1.80, 1.90, 1.94] \msol. These simulated measurements span a sufficiently wide range of the simulated hyperonic M-R curve, with two values corresponding to known masses for current NICER sources: PSR J1614-2230 (1.9 \msol, \citealt{Demorest2010}) and PSR J0437-4715 (1.4 \msol, \citealt{Reardon2016,Reardon2024}). The largest mass predicted in this injected data set is 1.94 \msol, a value that is still within the 95\% credible interval of the most massive pulsar and NICER source PSR J0740+6620 (which has a reported mass of 2.08$\pm 0.07$ \msol, \citealt{Cromartie2020,Fonseca_2021}).  This `Future-X' hyperon scenario is intended to illustrate the `best case' that we might be able to obtain with future improved X-ray telescope missions.   For the nucleonic injected data, as in \cite{Huang24}, we choose six measurements whose masses are centered at [1.20, 1.40, 1.90, 2.00, 2.10, 2.20] \msol. Three of these values correspond to known NICER sources, PSR J0740+6620, PSR J1614-2230 and PSR J0437-4715, with the other three chosen to span a wide range of masses. Again we assume a 2\% uncertainty. The two M-R curves, together with all simulated measurements, are shown in Figure \ref{fig:ground_truth}.

The two different injected M-R datasets with (inject-hyp) and without hyperons (inject-nucl) will be fitted by the two different models (nucleonic and hyperonic). The cross-comparison of these  four inferences will be performed to systematically test the ability of our future large area X-ray telescope data to discriminate the internal composition of neutron stars. We will compare model performance for the different sets of simulated observations using Bayes factors.

\section{EOS constraints from current observations}
\label{results_now}
 In this section we examine the constraints on EOS parameters and nuclear saturation quantities provided by all the current observations. Those are the M-R measurements derived from NICER observations for PSR J0030+0451 (\citealt{Riley2019}, $M= 1.34_{-0.16}^{+0.15}$ \msol~ and $R = 12. 71_{-1.19}^{+1.14}$ km) and PSR J0740+6620 (\citealt{Riley2021}, $M= 2.07\pm 0.07 $ \msol~and $R = 12.39_{-0.98}^{+1.30}$ km). Note that there are now newer results available for PSR J0030+0451 \citep{Vinciguerra24} and PSR J0740+6620 \citep{Salmi24} but since they are reasonably similar we do not update these.  We do not include any separate high mass measurements from radio pulsar timing since the current highest value - for PSR J0740+6620 - is included as a prior on the NICER M-R measurement for this source. We also include the two GW tidal deformability measurements GW170817 and GW190425 \citep{LIGOScientific:2017vwq,LIGOScientific:2020aai}. The chirp mass $M_{\text {c }}=\left(M_{1} M_{2}\right)^{3 / 5}/\left(M_{1}+M_{2}\right)^{1 / 5}$ is fixed to the median value $M_{\text {c1}}=1.186$ \msol~for GW170817 and $M_{\text {c2}}=1. 44$ \msol~for GW190425. 

\begin{figure*}
	\centering
	\includegraphics[scale=0.4]{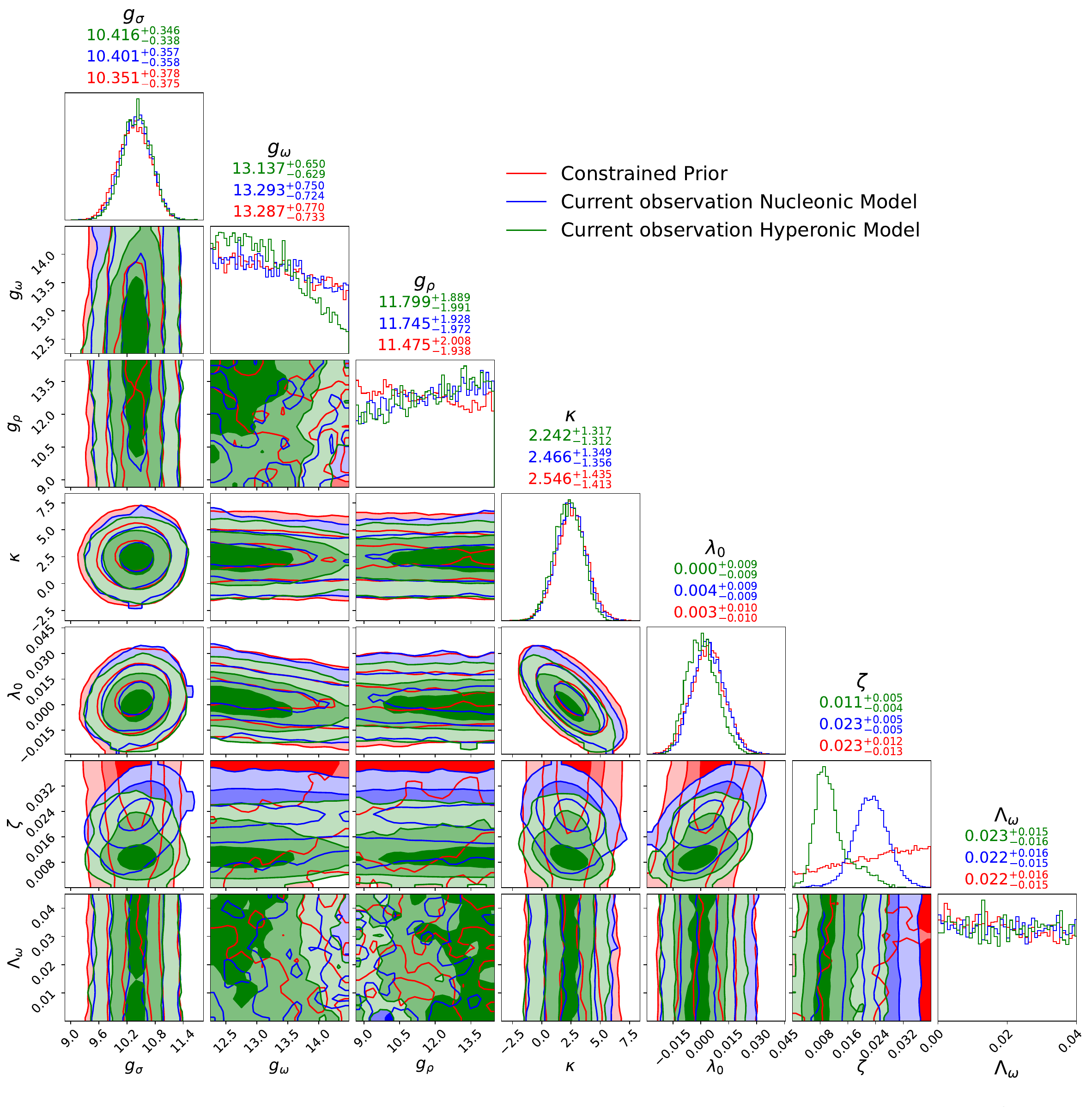}
	\caption{The posterior of the first seven nucleonic EOS model parameters derived from existing observations, for both pure nucleonic and hyperonic models. Blue is the inference result with the current observation but with the nucleonic model, green is the posterior of the same set of observations with the hyperonic model. The red contour is the constrained hyperonic prior space we used in these two analyses. The contour levels in the corner plot, going from deep to light colours, correspond to the 68\%, 84\% and 98.9\% levels. The title of each panel indicates the median of the distribution as well as the range of the 68\% credible interval. Here $\kappa$ is given in MeV.
	}
	\label{J0740}
\end{figure*}   

In Figure \ref{J0740}, we illustrate the posterior distribution resulting from the nucleonic and hyperonic models, as determined from current observations, compared to the constrained hyperonic prior defined in section \ref{prior}. We notice that after introducing the hyperonic part into the EOS to do the inference, the peak of $\zeta$ is shifted significantly to lower values. This is reasonable since the appearance of hyperons softens the EOS, but we still have to satisfy the observational constraints from the high-mass pulsar PSR J0740+6620. The inference naturally favours smaller values of $\zeta$ to stiffen the EOS, so as to predict stars with larger maximum masses and radii, as already seen in \cite{Huang24}. The reduction of the $\zeta$ parameter results in an increase of radius of a 1.4 \msol star. However, the $g_{\omega}$ and $\lambda_0$ values slightly decrease when using the hyperonic model, thus compensating the increase of the radii, so that our results for the radius are still compatible with measurements from PSR J0740+6620 and PSR J0030+0451, together with the tidal deformability determination from GW170817.

\begin{figure}
	\centering
	\includegraphics[width = \columnwidth]{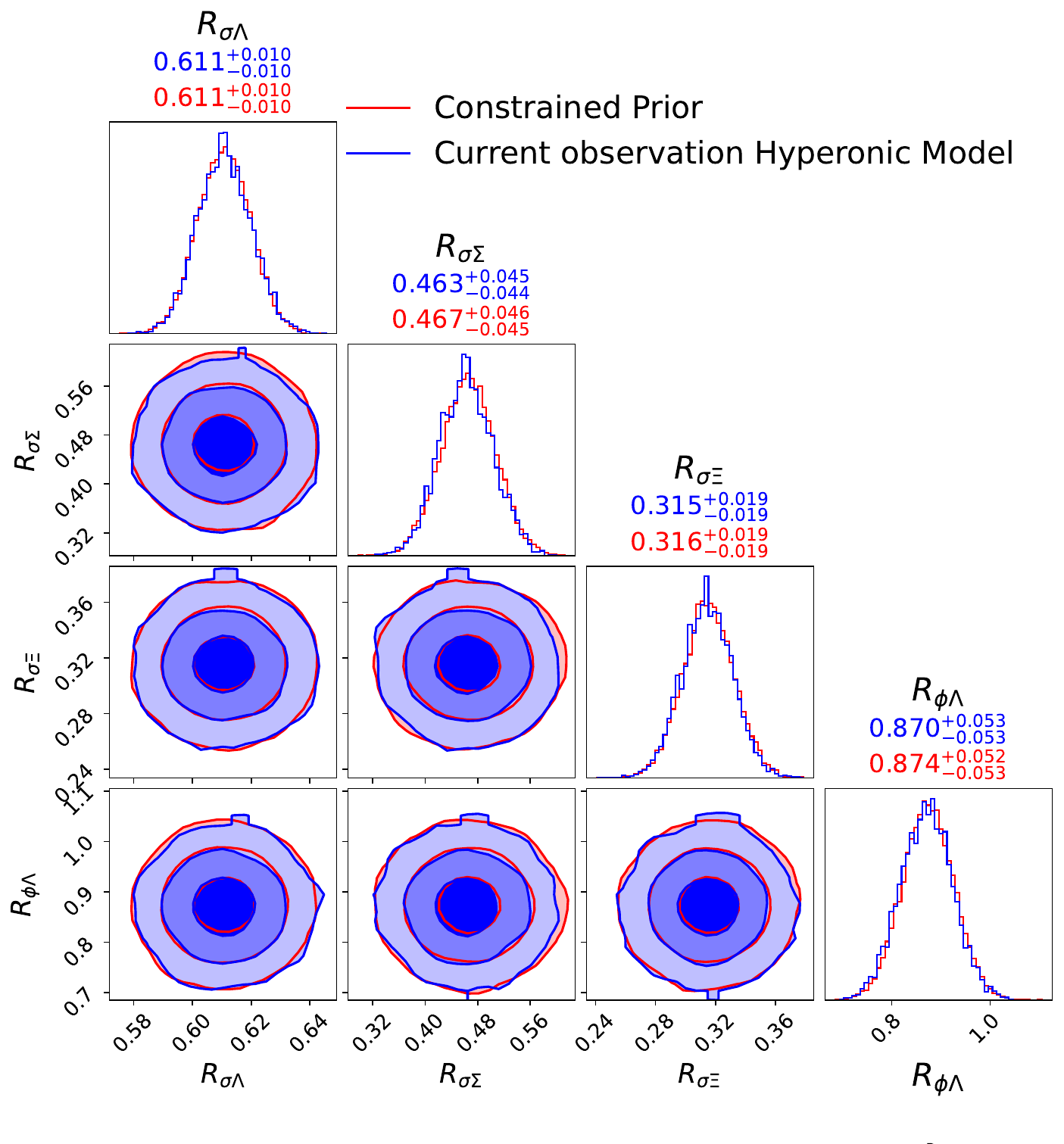}
	\caption{The posterior of the four hyperonic EOS model parameters after applying constraints from existing observations using the hyperonic prior. Red represents the prior of the hyperonic parameters. Blue indicates the inference result with the current observations (assuming the hyperonic model/prior). The contour levels in the corner plot, going from dark to light colours, correspond to the 68\%, 84\% and 98.9\% levels. The title of each panel indicates the median of the distribution as well as the range of the 68\% credible interval. 
 }
	\label{NICER_hypara}
\end{figure} 
\begin{figure}
	\centering
	\includegraphics[width = \columnwidth]{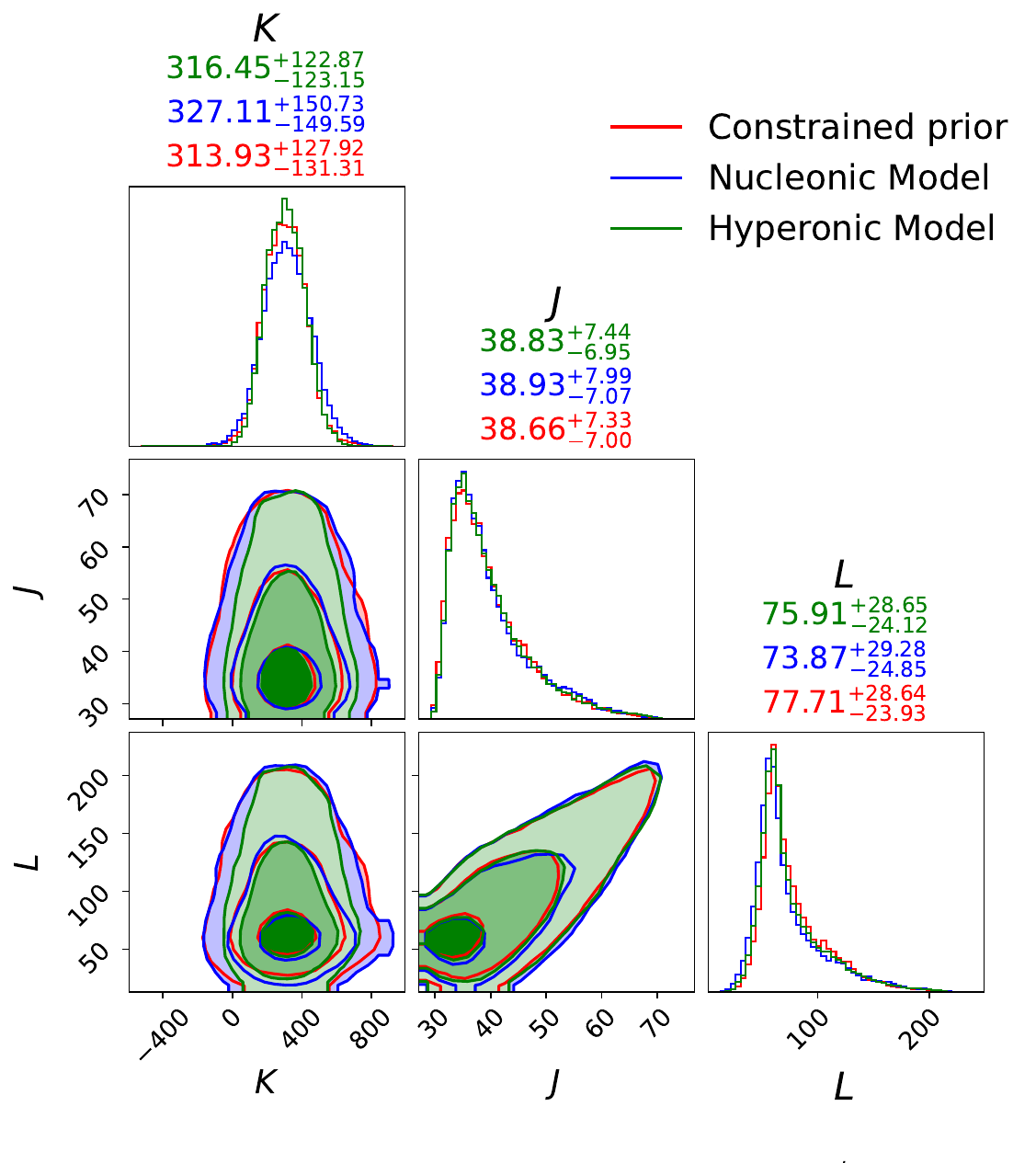}
	\caption{The posterior of all the nuclear quantities after applying constraints from current astrophysical observations using both the nucleonic model (blue contour) and the hyperonic model (green contour) for the inference. Red shows the constrained hyperonic prior. The contour levels in the corner plot, going from dark to light colours, correspond to the 68\%, 84\% and 98.9\% levels. The title of each panel indicates the median of the distribution as well as the range of the 68\% credible interval. }
	\label{Nuclear_2prior}
\end{figure} 
\begin{figure}
\includegraphics[width=\linewidth]{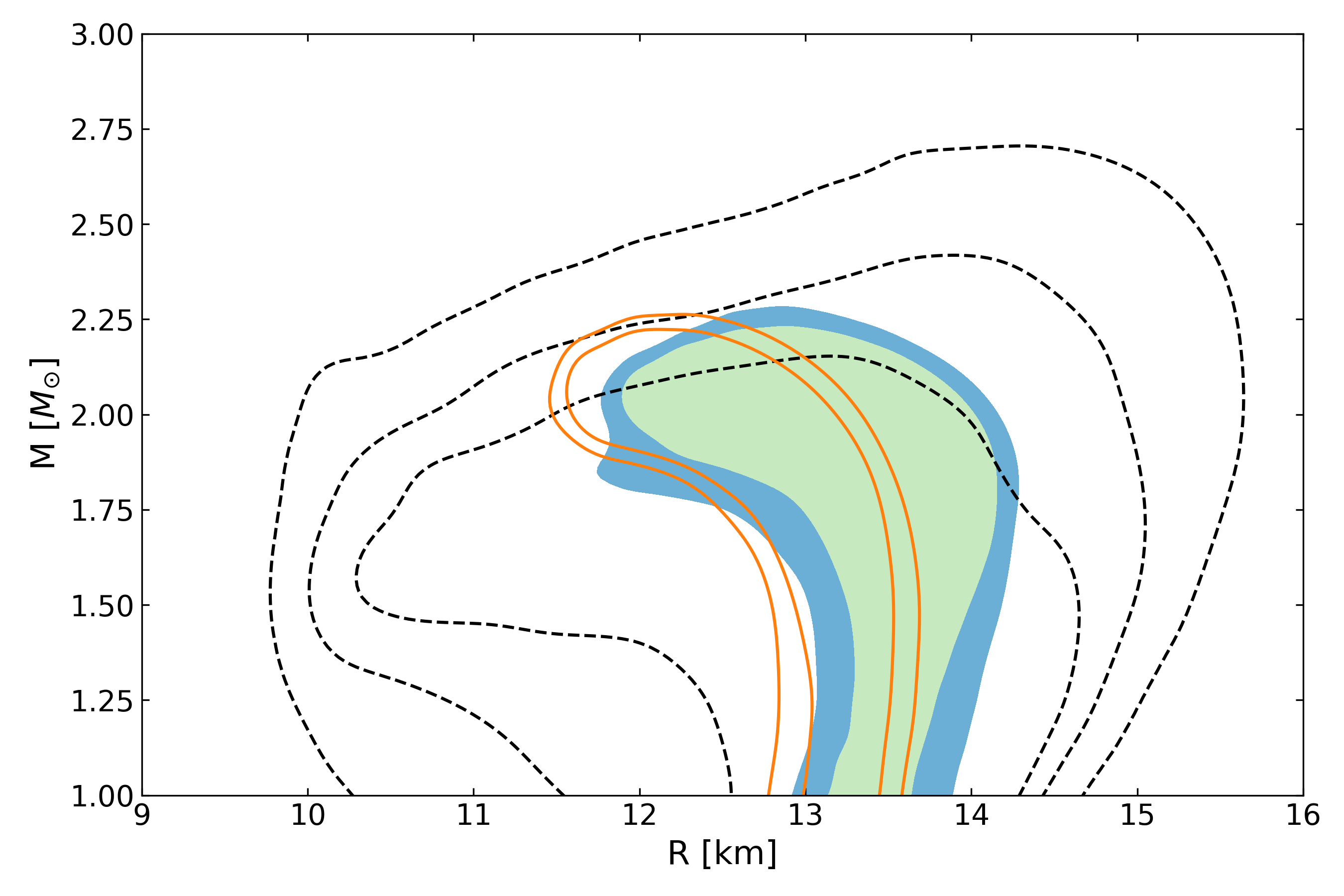}
\caption{The M-R posterior after applying all current observations (except PSR J0437-4715) using the hyperonic model, showing the 68\% (light green) and 84\% (blue) credible regions and the nucleonic posterior from \citep{Huang24} (orange line).   The black dashed lines show the M-R prior resulting from the initial hyperonic EOS model and priors but after imposing priors on $E/A$, $m^{*}$ and $n_0$, delineating the 68\%, 84\% and 100\% credible regions (as in Figure \ref{M-R_nu}).}
\label{compare-NICER} 
\end{figure}
The hyperonic parameters after inference using the current observations listed above are shown in Figure \ref{NICER_hypara}. It is clear that no significant constraint could be placed on the hyperonic parameters; current observations are still not sensitive to the neutron star composition and exotic degrees of freedom. This is in line with our conclusions in \cite{Huang24}. 

Figure \ref{Nuclear_2prior} shows the posterior of certain nuclear quantities, such as the incompressibility $K$, symmetry energy $J$ and slope of the symmetry energy $L$ constrained by current observation data, resulting from the hyperonic model inference and the nucleonic model one compared with the constrained hyperonic prior.  Inference using a hyperonic model as compared to a nucleonic model does not lead to any notable changes on the nuclear properties, except for a marginal reduction in the distribution width of K. This observation suggests that the current observational data are not yet sufficiently robust to induce significant shifts in the overall shape of the distributions, which is consistent with our previous analysis of the EOS parameters.
\begin{figure*}
	\centering
	\includegraphics[scale=0.4]{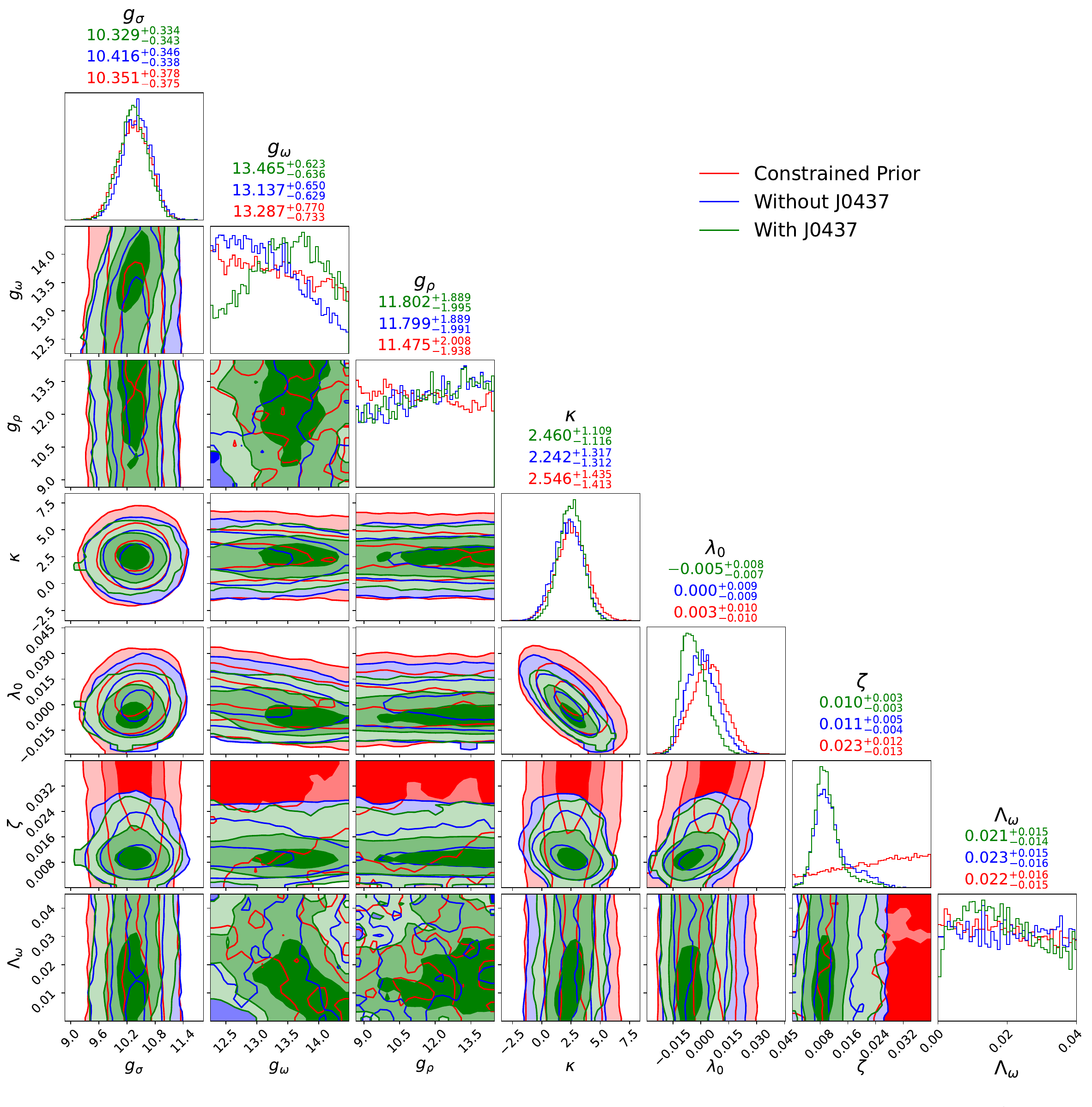}
	\caption{
Effect of including the mass-radius measurement of PSR J0437-4715  on the model parameters: a comparison between the constrained hyperonic prior (red contour) and the posteriors obtained without including (blue contour levels) and including (green contour) the constraints from PSR J0437-4715.
 The contour levels in the corner plot, going from deep to light colors, correspond to the 68\%, 84\%, and 98.9\% levels. The title of each panel indicates the median value of the distribution as well as the range of 68\% credible interval. Here $\kappa$ is given in MeV. 
	}
	\label{J0437_eos}
\end{figure*} 
\begin{figure}
	\centering
	\includegraphics[width = \columnwidth]{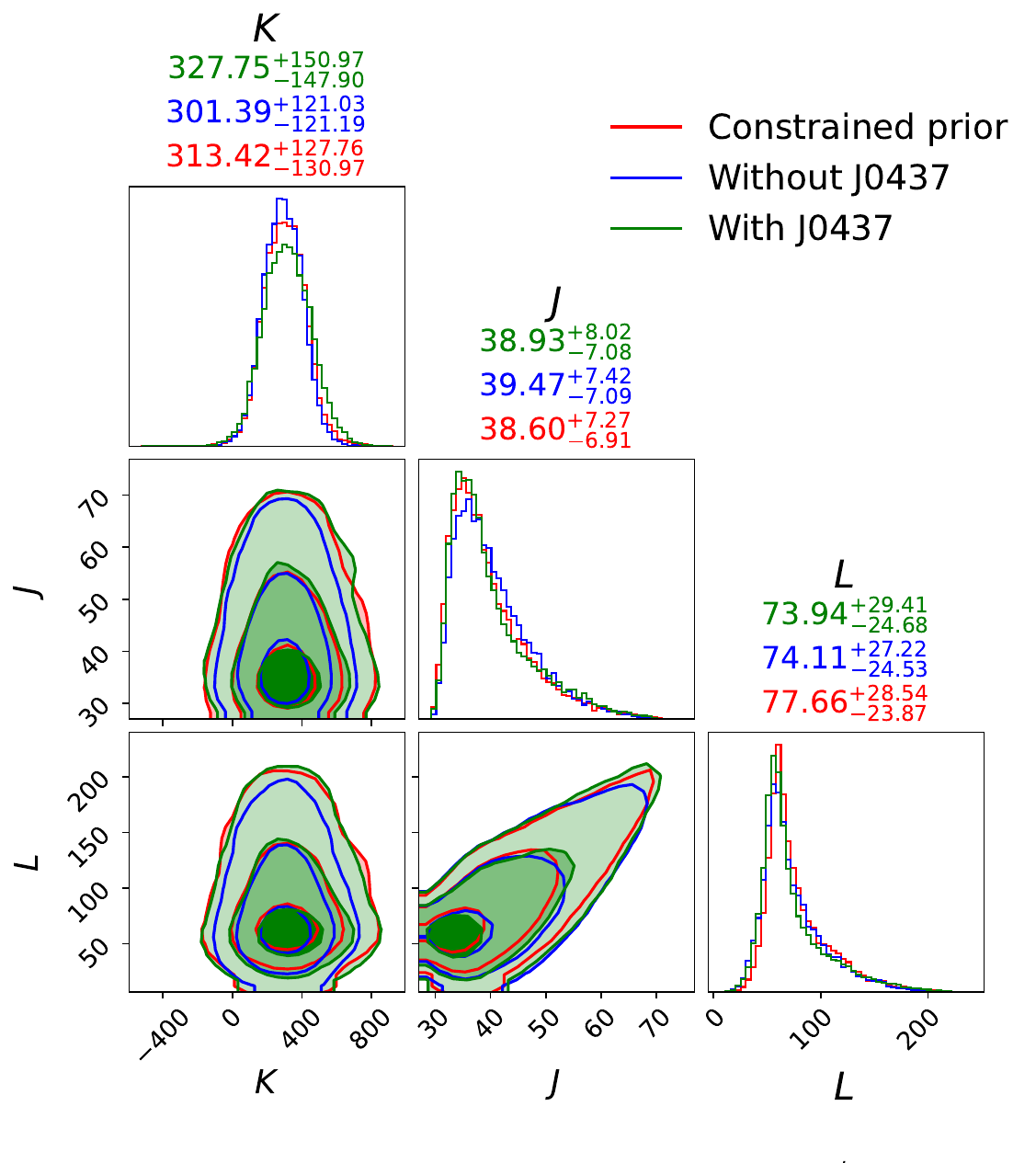}
	\caption{The posterior of all the nuclear quantities after applying constraints from current astrophysical observations with (green) and without (blue) the constraints from PSR J0437-4715. Red shows the constrained hyperonic prior. The contour levels in the corner plot, going from dark to light colours, correspond to the 68\%, 84\% and 98.9\% levels. The title of each panel indicates the median of the distribution as well as the range of the 68\% credible interval. }
	\label{Nuclear_j0437}
\end{figure} 
In Figure \ref{compare-NICER} we show the M-R posterior of the hyperonic EOS inference with all the current observational constraints. We also show the M-R constrained hyperonic prior. It is clear that the area of the allowed region has shrunk significantly after the current observational constraint when hyperons are considered. Compared with the M-R posterior of the nucleonic EOS inference in \citep{Huang24}, the posteriors have a similar area in the mass-radius plane, but the hyperonic M-R posterior is slightly wider. It is interesting to see that a consequence of including hyperons and imposing observational constraints, in particular from PSR J0740+6620, is a shift in the probability distributions towards larger radii for all neutron star masses. Similar conclusions were drawn in \citep{Malik:2022jqc,Malik:2023mnx}. This reflects the fact that the symmetric nuclear matter EOS has to be much harder to meet the high mass constraint.

The constraining power of astrophysical observations on hyperon-meson couplings has also been the subject of discussion in \citep{Sun22}.  In this study, the authors constrained the couplings of the $\sigma$ and $\omega$ mesons to the $\Lambda$ hyperon within a Bayesian inference procedure, imposing separate astrophysical and hypernuclear constraints and considering a fixed set of nuclear EOS. They start from a uniform distribution for the ratios $R_{\sigma\Lambda}=g_{\sigma\Lambda}/g_{\sigma N}$ and $R_{\omega\Lambda}=g_{\omega\Lambda}/g_{\omega N}$, assuming that they should be less than 1. They verify that the hypernuclear data are the most constraining. This is consistent with our results showing that astrophysical observations do not constrain the hyperon-meson couplings, beyond the information we already have from hypernuclei.

\subsection{Constraint from PSR J0437-4715}
\label{j0437}
Recently \citet{Choudhury24} reported the results of pulse-profile modeling using NICER data of PSR J0437-4715 (J0437), the closest and brightest rotation-powered millisecond pulsar. Using a mass prior from radio timing \citep{Reardon2024} they reported a mass of $M=1.418 \pm 0.037 \mathrm{M}_{\odot}$ and a radius of $R=11.36_{-0.63}^{+0.95} \mathrm{~km}$ (68\% credible intervals).  This source has a mass very close to that inferred for PSR J0030+0451, while the radius is slightly different. It could therefore be a very good source for constraining a phase transition or the onset of a new degree of freedom such as hyperons. 

In Figure \ref{J0437_eos} we compare the posteriors with and without the J0437 constraint. The inclusion of the new data favors the upper end of the prior for $g_{\omega}$, while $\lambda_0$ shows a narrower 68\% interval shifted towards smaller values and the credible interval on the $\zeta$ parameter also narrows. The behaviour obtained for $\lambda_0$ is consistent with our expectation that a smaller value corresponds to a smaller
radius of the 1.4 $M_{\odot}$ star. The reduction of $\lambda_0$ comes with an increase of $g_{\omega}$ and a reduction  of $\zeta $ to make the model stiff enough to also describe the two solar mass pulsar PSR J0740+6620.  It is interesting to note that the parameter $\Lambda_{\omega}$ begins to take a shape different from the prior distribution, highlighting the increased constraining power of the new data. 

Regarding the hyperonic parameters, the current precision of this new result still does not allow us to draw any obvious constraints. In Figure \ref{Nuclear_j0437}, we present the constrained nuclear quantities, comparing scenarios with and without J0437. There is no clear difference, leading to the conclusion that even with this new observation, we cannot constrain nuclear quantities in this hyperonic model.

\section{EOS constraints from future observations}
\label{results_future}
In this section, we explore the extent to which future telescopes will be able to detect the composition of neutron stars. We do this by computing the evidence for each inference run, thereby allowing model comparison based on the same injected data. Our overall goal is to see whether we can accurately recover the models that generated the injected data and whether we can discriminate between the different models. Specifically, we want to determine whether our simulated observational data preferentially support the existence of hyperons in the interior of neutron stars. 
\begin{figure*}
	\centering
	\includegraphics[scale=0.4]{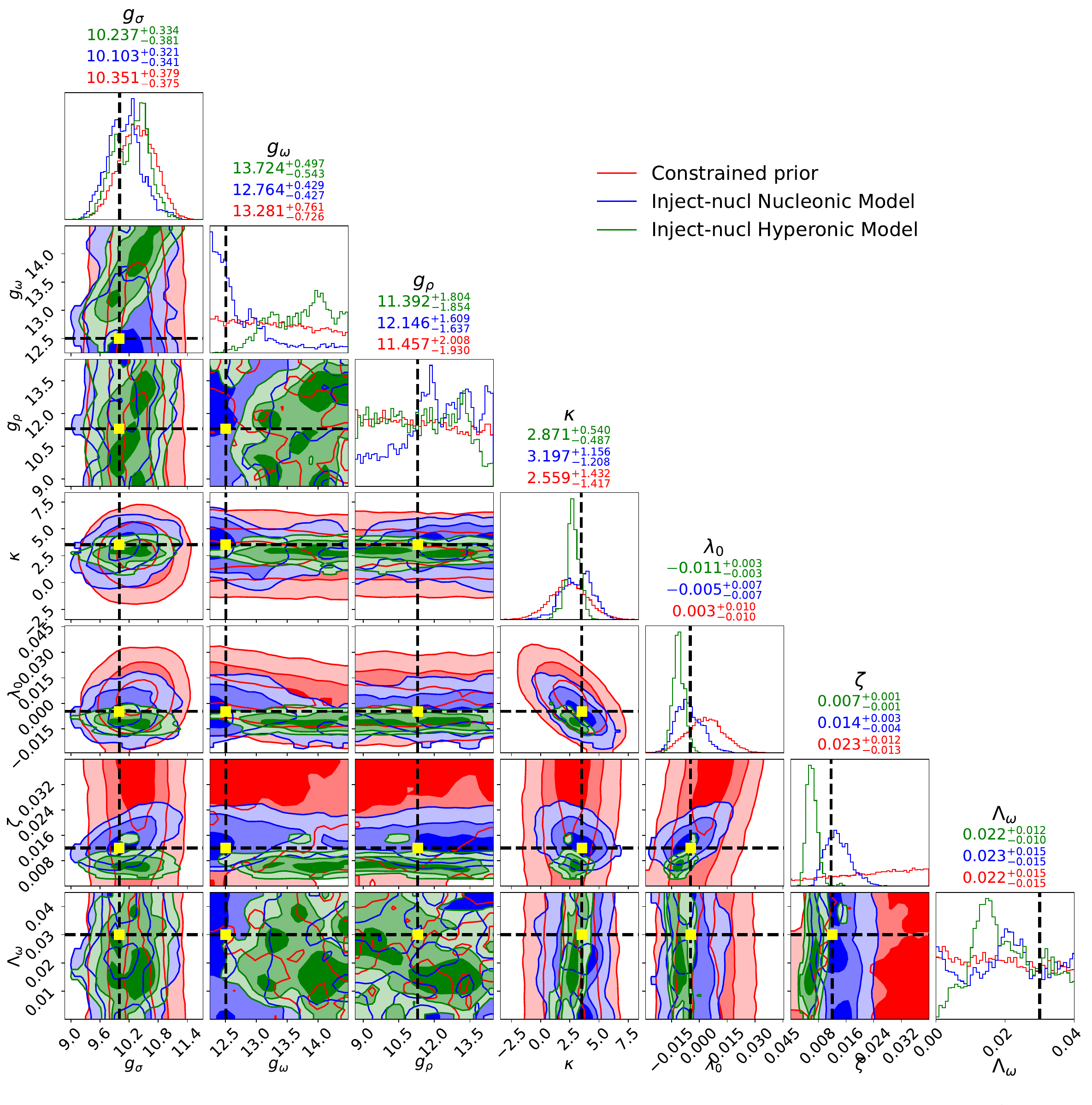}
	\caption{After applying the constraint of simulated Future-X inject-nucl measurements, i.e. six 2\% uncertainty M-R measurements along the M-R curve obtained with the  TM1-2$\omega\rho$n EOS (nucleonic EOS), the posterior of all EOS parameters from the nucleonic model (blue contour levels) is compared with the results from the hyperonic model (green contour) and the constrained hyperonic prior (red contour). The contour levels in the corner plot, going from deep to light colours, correspond to the 68\%, 84\% and 98.9\% levels. The title of each panel indicates the median of the distribution as well as the range of the 68\% credible interval. Here $\kappa$ is given in MeV. The black dashed horizontal and vertical lines in the plot and the yellow dots show the values for the injected EOS used to generate the simulated M-R measurements.  
	}
	\label{Nu_two_prior}
\end{figure*} 
\begin{figure}
	\centering	\includegraphics[scale=0.35]{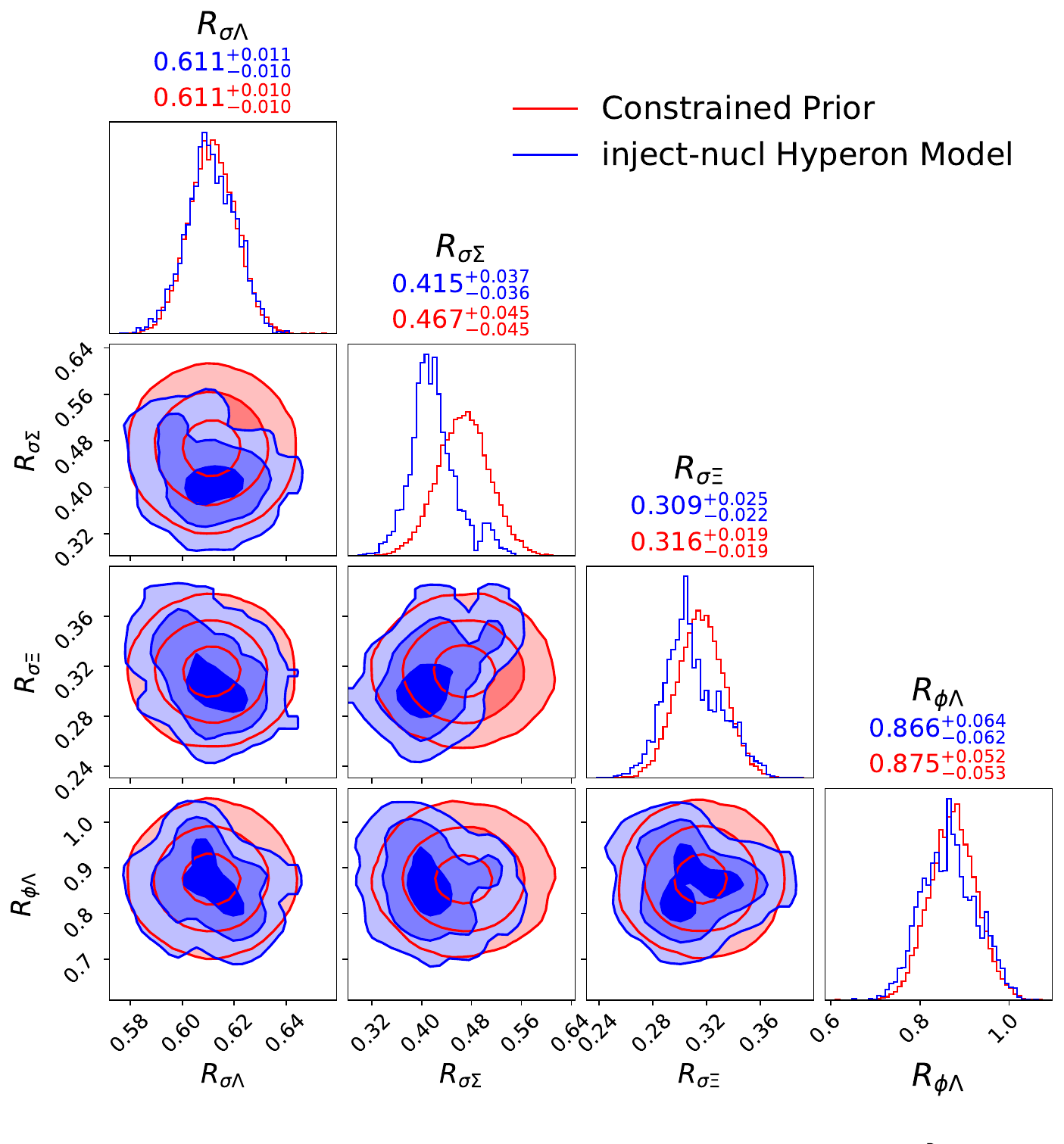}
	\caption{The posterior of the four hyperonic EOS model parameters after applying constraints from the Future-X inject-nucl data using the hyperonic model (blue). Red represents the priors for the hyperonic parameters. The contour levels in the corner plot, going from dark to light colours, correspond to the 68\%, 84\% and 98.9\% levels. The title of each panel indicates the median of the distribution as well as the range of the 68\% credible interval.
	}
	\label{hyperon-para_f-NGT}
\end{figure} 
\begin{figure}
	\centering
	\includegraphics[scale=0.4]{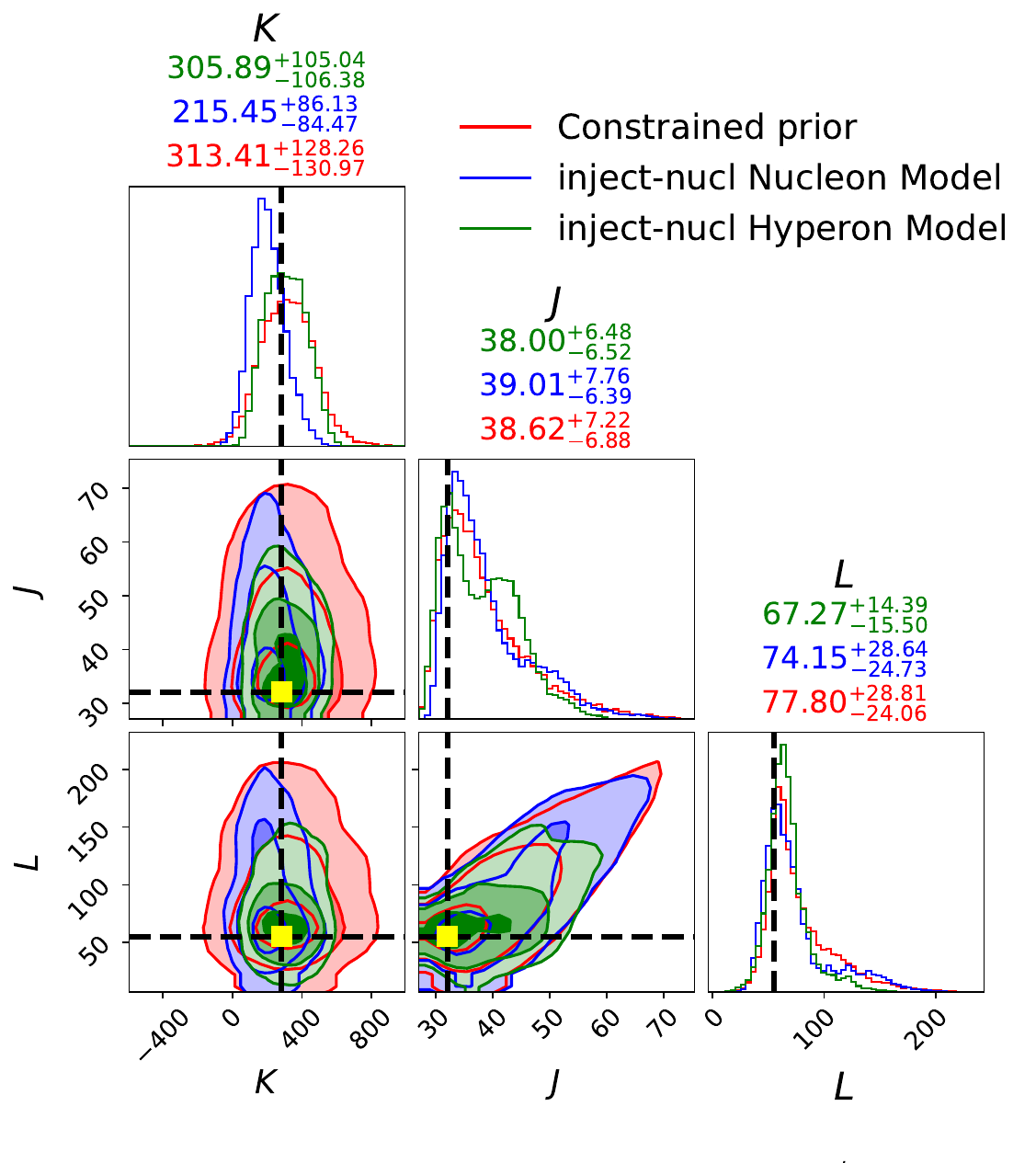}
	\caption{The posterior distributions of the nuclear quantities obtained imposing the Future-X inject-nucl data from TM1-2$\omega\rho$n. Red shows the constrained hyperonic prior, blue and green show, respectively, the posterior resulting from the nucleonic model and the hyperonic model. The contour levels in the corner plots, going from dark to light colours, correspond to the 68\%, 84\% and 98.9\% levels. The title of of each panel indicates the median of the distribution as well as the range of the 68\% credible interval. The black dashed horizontal and vertical lines in the plot and the yellow dots show the injected values used to generate the simulated M-R measurements. 
	}
	\label{Nuclear_f-NGT}
\end{figure} 

The following subsections will be organized according to the different injected data, each of the injected data reflecting the two types of "reality": neutron stars with hyperons (inject-hyp) and neutron stars without hyperons (inject-nucl).

\subsection{Nucleonic injected data}
In this subsection, we investigate the performance of the models under the nucleonic injected data, inject-nucl. 

For the Future-X inject-nucl data, Figure \ref{Nu_two_prior} shows the posterior of the nucleonic EOS parameters using the nucleonic model and the hyperonic model. The first general observation is that the same set of EOS parameters is modified with both prior settings, namely $g_{\omega}$, $g_{\rho}$, $\lambda_0$, $\zeta$ and $\lambda_{\omega}$. In addition, with the hyperonic model there is a tighter constraint on these nucleonic parameters, since satisfying the observational constraints simultaneously with allowing the existence of hyperons limits the nucleonic EOS parameter space more tightly.

Similar to the analysis performed for the current observations,  the Future-X inject-nucl constraints favour smaller $\zeta$ in order to reach the largest mass, which now reaches 2.20 \msol, and  $\lambda_0$ and $g_{\omega}$ reshape accordingly to predict a reasonable radius for 1.4 \msol star.  However, in contrast to the current observation scenario, all EOS parameters shift compared to their priors, indicating the potential for strong EOS constraints from next generation X-ray telescopes. The consistency of the inferred posteriors with the injected parameter values is evident, indicating an effective approach to model recovery. In contrast, the posterior inferred  from hyperonic model tend to converge to a distinct region. This discrepancy is due to the inclusion of hyperons, which requires adjustments to the nucleonic model parameters. 
\begin{figure}
	\centering
	\includegraphics[scale=0.35]{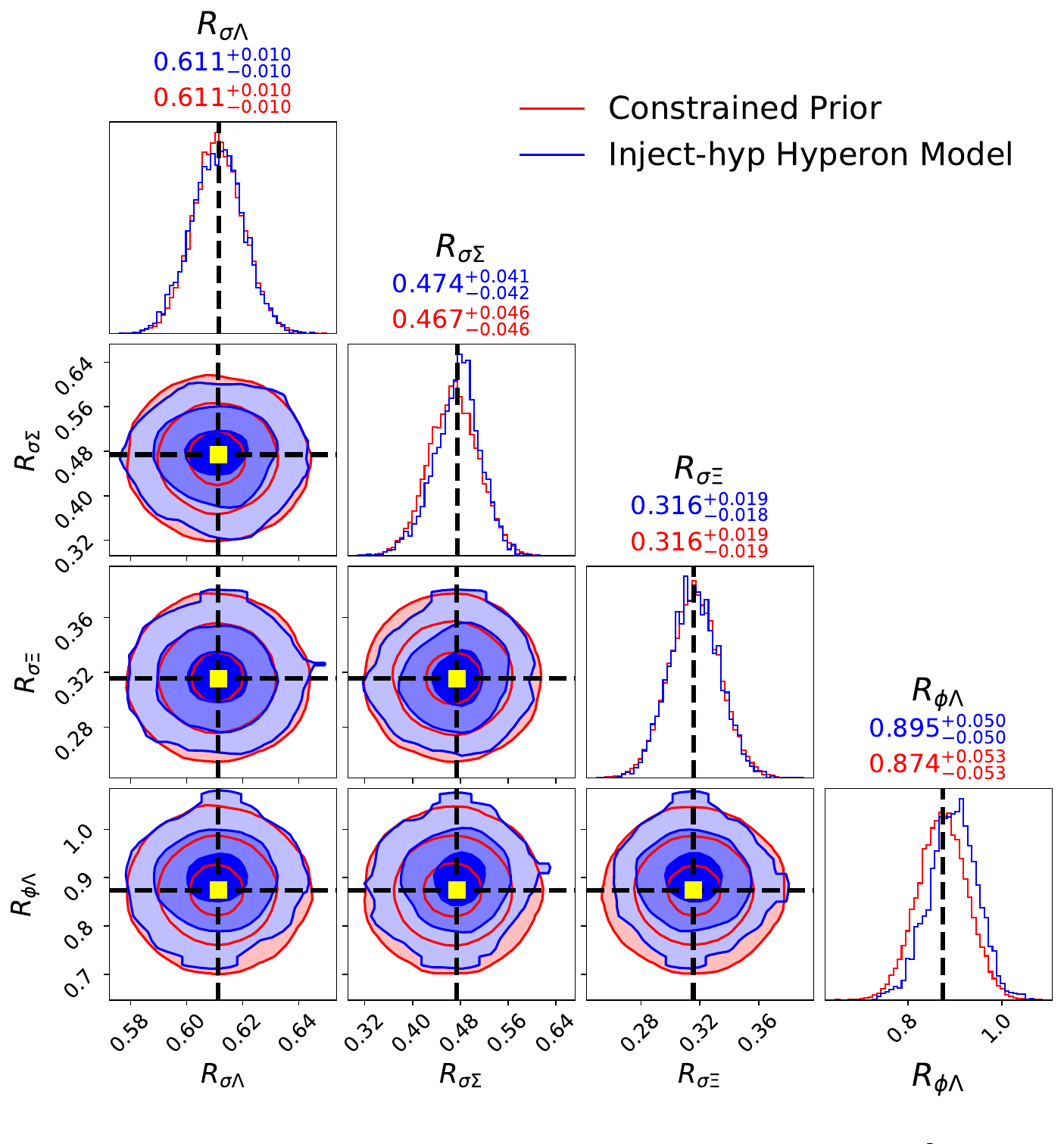}
	\caption{The posterior of the four hyperonic EOS model parameters after applying the constraints from the Future-X hyperon case, inject-hyp data, using the hyperonic model (blue). Red represents the prior of the hyperonic parameters in the hyperonic model. The contour levels in the corner plot, going from deep to light colors, correspond to the 68\%, 84\%, and 98.9\% levels. The title of each panel indicates the median value of the distribution as well as the range of 68\% credible interval. The black dashed horizontal and vertical lines in the plot and the yellow dots show the injected values used to generate the simulated M-R measurements.
	}
	\label{hyperon-para_f-HGT}
\end{figure} 
\begin{figure*}
	\centering
	\includegraphics[scale=0.4]{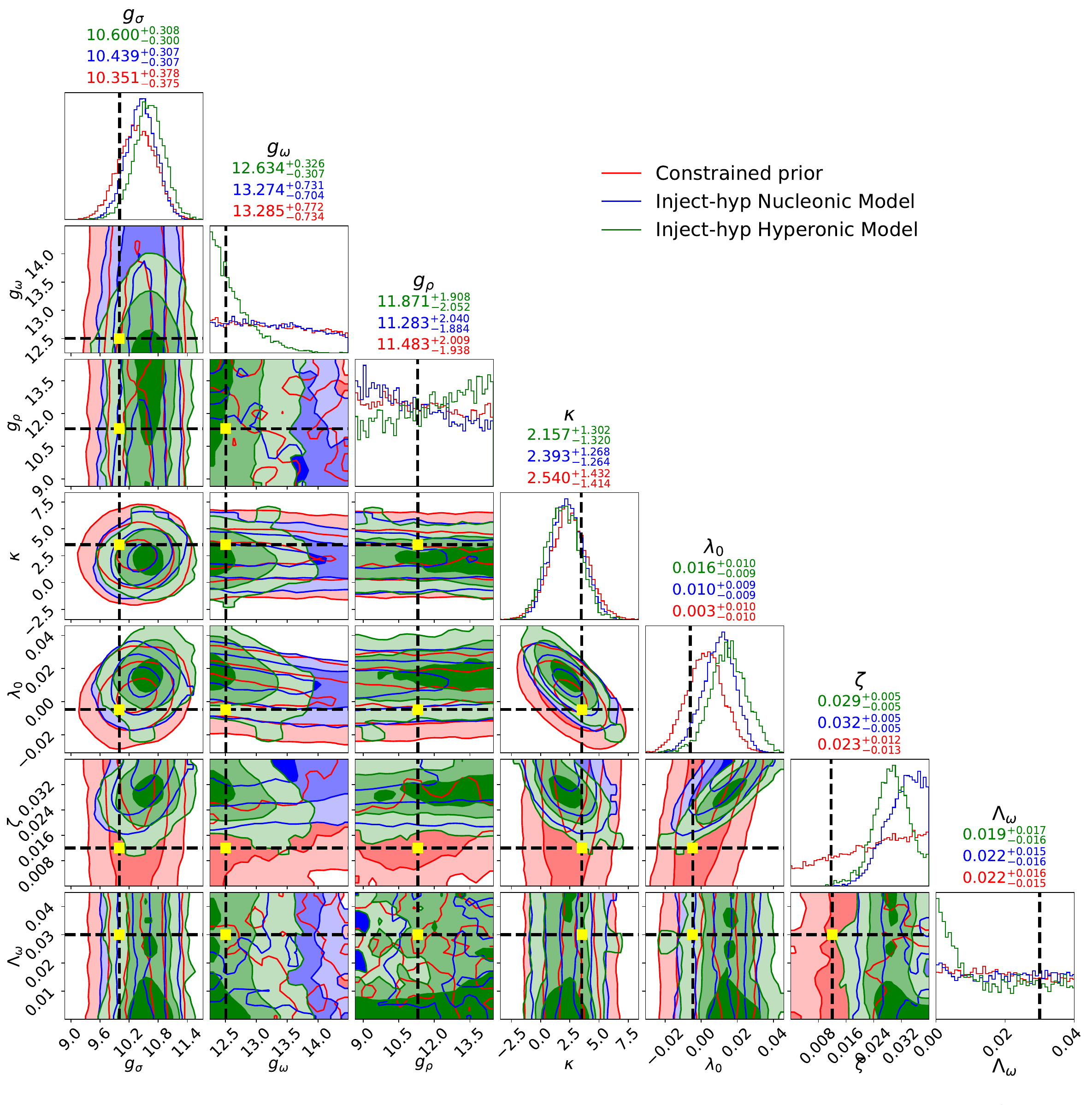}
	\caption{After applying the constraint of simulated Future-X inject-hyp measurements, that is six 2\% uncertainty M-R measurements along M-R curve obtained with the TM1-2$\omega\rho$nH EOS (Hyperonic EOS), the posterior of all the EOS parameters from nucleonic model (blue contour levels) compared to the results from hyperonic model (green contour) and the constrained hyperonic prior (red contour). The contour levels in the corner plot, going from deep to light colors, correspond to the 68\%, 84\%, and 98.9\% levels. The title of each panel indicates the median value of the distribution as well as the range of 68\% credible interval. Here $\kappa$ is given in MeV. The black dashed horizontal and vertical lines in the plot and the yellow dots show the injected values used to generate the simulated M-R measurements. 
	}
	\label{hynu_compare}
\end{figure*} 
\begin{figure}
	\centering
	\includegraphics[scale=0.35]{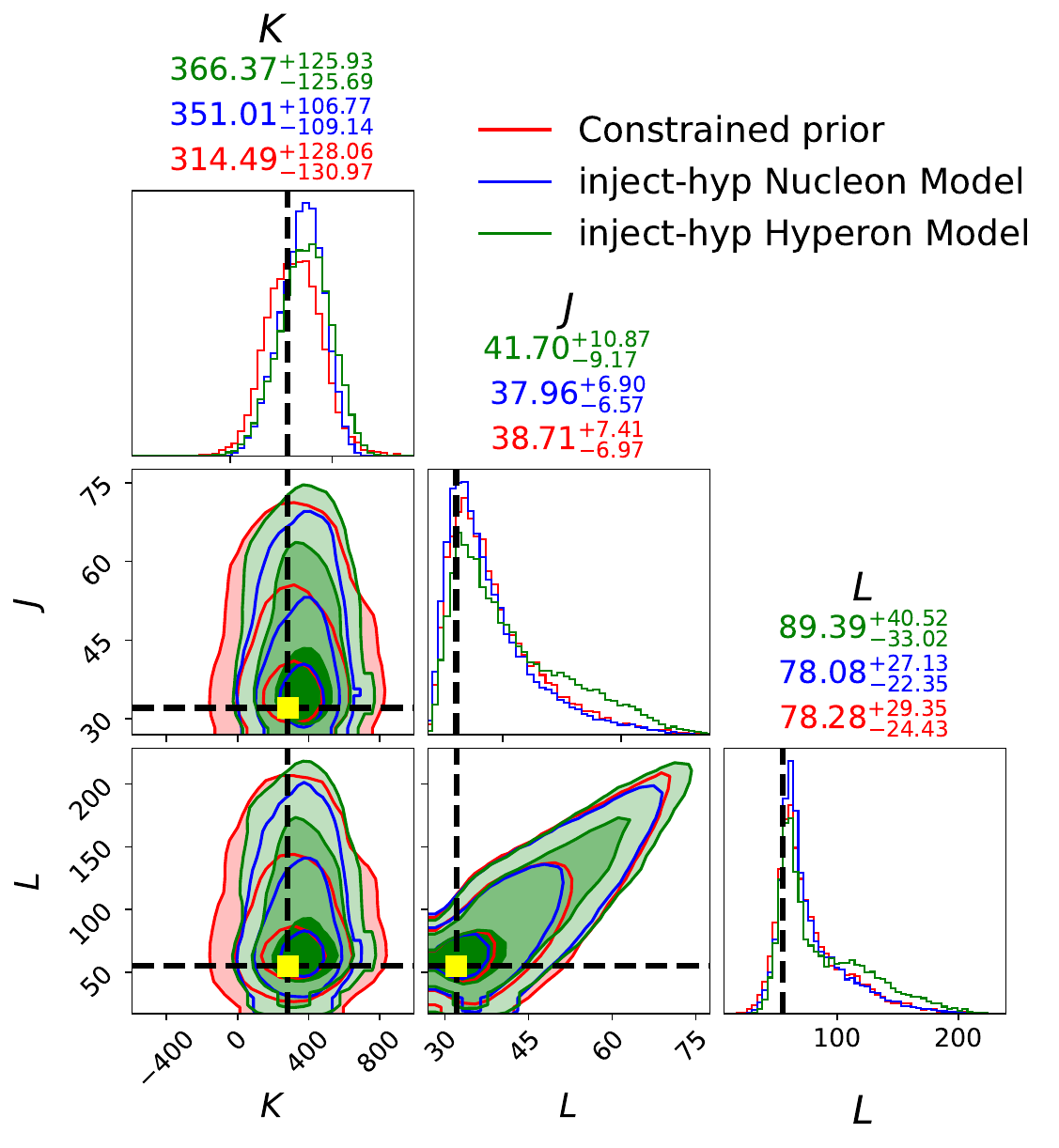}
	\caption{The posterior of all the nuclear quantities after applying constraints of simulated Future-X inject-hyp measurements for both nucleonic model (blue contour) and hyperonic model (green contour). Red shows the constrained hyperonic prior. The contour levels in the corner plot, going from deep to light colors, correspond to the 68\%, 84\%, and 98.9\% levels.  The title of each panel indicates the median value of the distribution as well as the range of 68\% credible interval. }
	\label{nuclear-f-HGT}
\end{figure} 

In Figure \ref{hyperon-para_f-NGT} we illustrate the constraints imposed on the hyperonic parameters by the nucleonic injected data. The hyperonic parameters appear to more restricted. However, it is crucial not to interpret this as favoring hyperons in the simulated sources, since the EOS that generated the injected data does not include hyperons. Instead, we should interpret these constraints as indicating a reduction in the hyperonic parameter space due to neutron star measurements, and a shift to a more repulsive interaction. As the measurements continue to improve, it is foreseeable that the hyperonic parameter space will become increasingly constrained, potentially posing challenges in finding suitable parameter values to effectively fit the observations. This underlines the evolving nature of our understanding as observational capabilities improve. 

In Figure \ref{Nuclear_f-NGT} we illustrate the constraints on the selected nuclear quantities.  By comparing  the different model constraints, we observe that the hyperonic model imposes tighter constraints on the nuclear properties. This observation is consistent with the constraining power seen in the EOS space shown in Figure \ref{Nu_two_prior}. It seems that to account for the presence of hyperons, finer adjustments to nuclear quantities are required to mitigate the softening effects on the EOS induced by hyperons.

\subsection{Hyperonic injected data}

In this subsection, we delve into the constraints introduced by the Future-X hyperon scenario, inject-hyp data. 

Figure \ref{hyperon-para_f-HGT} illustrates the constraints introduced by the inject-hyp data obtained from TM1-2$\omega\rho$nH. Since the injected EOS predicts a maximum mass neutron star of 1.94 \msol, this places less stringent constraints on hyperons, since hyperons are less likely to exist if large-mass neutron stars are observed. As a result, the posterior shows minimal reshaping, even with a 2\% uncertainty in the precision of the mass-radius measurement. 

Turning to the nucleonic EOS parameters shown in Figure \ref{hynu_compare}, several notable observations can be made. Firstly, the nucleonic parameters tend to be better constrained when the hyperonic model is used, particularly $g_{\omega}$ and $\Lambda_{\omega}$. In the case of the nucleonic model, the posteriors on these parameters are not significantly different from the prior distribution. 

Another important observation concerns the sensitivity of $\zeta$ to the maximum mass of neutron stars. In the nucleonic model, a large $\zeta$ is required to soften the EOS if the maximum mass and its radius are not large, such as the 1.94 \msol star of the injected hyp observations (see the discussion of the effect of $\zeta$ in \citealt{Fattoyev:2010mx}). In an hyperonic model, the appearance of hyperons has an effect on the EOS and the M-R relation similar to the effect of $\zeta$; it softens the EOS at high densities giving rise to smaller maximum masses with smaller radii. As a consequence, the hyperonic model does not predict values of $\zeta$ as large as the nucleonic model because the role of the term with the coupling $\zeta$  is partially overtaken by the hyperon onset. 

When comparing the recovery of the injected EOS parameters (indicated by the black dashed line), for $\zeta$, the injected value lies outside the 3-sigma range of nucleonic posterior. This observation suggests that the underlying hyperonic EOS is finely-tuned with specific nucleonic and hyperonic parameter choices, making it challenging to recover through inference and necessitating more precise measurements or more measurements.

The resulting distribution of the selected nuclear quantities is shown in Figure~\ref{nuclear-f-HGT}. Interestingly, this figure shows that the hyperonic posterior is less constrained than the nucleonic posterior. However, it is important to note that these constraints are still not very tight.

\subsection{Bayesian Evidence Comparison}

\begin{table}
\centering

\begin{tabular}{ccc}
\hline \hline 
\text{Injected data + Model}  &\text{ln(Z)} & \text{Bayes' Factor}\\ 
\hline 
\text{Nucleonic + Nucleonic}     & -210.05      & \text{...}\\
\text{Nucleonic + Hyperonic} &  -211.85     & \text{N/H = 6.04}\\
\hline
\text{Hyperonic + Nucleonic}  & -116.68 & \text{...}   \\ 
\text{Hyperonic + Hyperonic} & -111.45  & \text{H/N = 186.79}\\
\hline \hline
\end{tabular}

\caption{This table gives the global log evidence ($\ln Z$), as returned by Ultranest, for the nucleonic and hyperonic models under two different injected model simulated observations: inject-nucl and inject-hyp.  For comparison, we also give the Bayes' factor of these two different groups based on the same injected model, N/H denotes the Bayes' factor of the nucleonic vs. hyperonic model. H/N denotes the hyperonic model compared to the nucleonic one. 
}
\label{Bayes_factor}
\end{table}

In order to compare how well different models can recover the underlying injected EOS, we use Bayes' factors. In \citet{Kass} a model is deemed `substantially preferred' if the Bayes factor is greater than 3.2 and `strongly preferred' if it is greater than 10, and `decisive' if the Bayes' factor is greater than 100.

When calculating the Bayes' factor, we extract the logarithm of the Bayesian evidence directly from the inference, denoted as $\ln(Z)$. Here we focus exclusively on the comparison of the Future-X case study. The ultimate goal of the Bayesian inference of the EOS is to use Bayesian techniques to uncover the underlying EOS of neutron star matter or to extract information about neutron star composition.

In this study, the injected EOS act as the "hypothesised" underlying EOS of neutron stars. Our goal is to use different neutron star models (nucleonic or hyperonic) to recover the underlying EOS on which all simulated observations are based. By comparing the Bayesian evidence and then calculating the Bayes factor for each pair of inferences, we can gain unique insights into which model is preferred by the observations. This approach has significant potential for exploring exotic degrees of freedom.

In Table \ref{Bayes_factor}, we present the $\ln(Z)$ values along with the Bayes factors based on the performance with the different injected EOS. For the injected EOS that is purely nucleonic, the Bayesian evidence suggests that the nucleonic model performs better in explaining all simulated observations. This result is in line with the expectations, given that the underlying injected EOS is exclusively nucleonic. If we compute the Bayes factor between these two inferences to compare their fitting degree, we obtain a Bayes factor of 6.04. According to the interpretation of the Bayes factor we can confidently say that our simulated observations substantially favour the nucleonic model.

If the underlying injected EOS contains exotic degrees of freedom, such as hyperons, we compute the $\ln(Z)$ values using both nucleonic and hyperonic model frameworks. It is expected that the hyperonic model will exhibit significantly higher $\ln(Z)$ performance in recovering the hyperonic EOS, as the nucleonic model does not account for any exotic degrees of freedom. The resulting Bayes factor between these two inferences is 186.76. Despite the limitations of our study this is encouraging; even though we cannot fully recover the underlying hyperonic parameters, the presence of hyperons is still (correctly, given our injected data set) supported by the evidence.

\section{Conclusions}

We have considered a microscopic phenomenological hadronic model based on a field-theoretical framework, which includes both nucleonic and hyperonic degrees of freedom to span the neutron star M-R space. Using a Bayesian inference approach, the parameters of the model are constrained using observational data of neutron stars and some minimal information on the saturation properties of nuclear matter at saturation density. Our main objective was to complement the study \cite{Huang24} and to understand the capacity of neutron star observations to give information about neutron star interior composition. In \cite{Huang24} only nucleonic degrees of freedom were considered. By comparing the hyperonic model inference results calculated in the present study with the nucleonic model results built in \cite{Huang24}, we wanted to understand if it is possible to make a statement about the presence of hyperons in neutron stars.

As constraints, we consider two possibilities: i) the current observational constraints, including electromagnetic radio and X-ray and GWs; ii) simulated observations as expected from future X-ray timing telescopes, i.e. assuming that the neutron star radius and mass are determined with 2\% uncertainty. The simulated observations consist of two sets of six mass-radius observations with 2\% uncertainty, centered on the M-R predicted by two different EOS, one including only nucleons and leptons as degrees of freedom, and a second including also hyperonic degrees of freedom .

It was shown that current observations do not place a significant constraint on the hyperonic parameters of the model. A similar conclusion was reached in \cite{Sun22}. Regarding other model parameters, it has been shown that the hyperonic inference model favours parameters values that stiffen the EOS to predict stars with larger maximum masses and radii. However, the main conclusion  is that the hyperonic inference model constrained by current observations does not lead to significant changes in the nuclear properties.

In a second step we have considered simulated future X-ray data to constrain the hyperonic  and the nucleonic inference models. Each one of the models was constrained by simulated observational data generated from an EOS with and without hyperonic degrees of freedom. 

The nucleonic injected data placed noticeable constraints on the nucleonic EOS parameters of both inference models, with tighter constraints occurring for the hyperonic inference model due to the difficulty of simultaneously allowing the existence of hyperons and reaching the largest observed mass. The nucleonic injected data also had a non-negligible effect on the hyperonic parameters, showing an effective reduction of the hyperonic parameter space due to neutron star measurements. If no hyperons occur inside neutron stars, we may expect that tighter observational constraints could shift the hyperonic parameter space to make the interaction more repulsive. Finally, it was also shown that the hyperonic model imposes tighter constraints on the nuclear properties. 

Imposing as constraints the hyperonic injected data, which includes a maximum mass 1.94$M_\odot$ star, the hyperonic parameter space showed only a  minimal reshaping. However, the nucleonic EOS parameters of the hyperonic inference model were, in general,  better constrained than those of the nucleonic inference model. A parameter that was quite sensitive to the constraints is the parameter $\zeta$ that controls the behavior of the EOS at high densities, a larger value corresponding to a softer EOS and a smaller maximum mass, larger values being obtained with the nucleonic inference model. The effect of this parameter on the M-R curve is similar to that due to the presence of hyperons, and the two effects cannot be easily disentangled. For the injected data considered, the  $\zeta$ parameter of the injected EOS could not be recovered well (lying outside the 3-sigma posterior distribution of nucleonic inference model but still inside the 3-sigma distribution of the hyperonic parameters posterior), indicating that more precise - and possibly more - data are necessary.

Finally, we also compared the different future X-ray data scenarios by calculating Bayes' factors from the evidence of the two inference models when both sets of injected data (nucleonic and hyperonic) were considered as constraints. Using injected nucleonic data, the nucleonic inference came out as preferred; while when using injected hyperonic data, the Bayes' factor decisively favours the hyperonic inference model.  This result allows us to conclude that it is possible to obtain decisive evidence in favour of the hyperonic inference model in this case, even though we are not able to recover the EOS parameters.

\section*{Acknowledgements}

C.H. acknowledges support from an Arts \& Sciences Fellowship at Washington University in St. Louis and NASA grant 80NSSC24K1095. L.T. work was supported under contract No.~PID2022-139427NB-I00 financed by the Spanish MCIN/AEI/10.13039/501100011033/FEDER,UE and from the project CEX2020-001058-M Unidad de Excelencia ``Mar\'{\i}a de Maeztu''. Moreover, this project has received funding from the European Union Horizon 2020 research and innovation programme under the program H2020-INFRAIA-2018-1, grant agreement No.\,824093 of the STRONG-2020 project, from the CRC-TR 211 'Strong-interaction matter under extreme conditions'- project Nr. 315477589 - TRR 211, from the Generalitat de Catalunya under contract 2021 SGR 00171 and
Generalitat Valenciana under contract CIPROM/2023/59.  C.P. received support from Fundação para a Ciência e a Tecnologia (FCT), I.P., Portugal, under the  projects UIDB/04564/2020 (doi:10.54499/UIDB/04564/2020), UIDP/04564/2020 (doi:10.54499/UIDP/04564/2020), and 2022.06460.PTDC (doi:10.54499/2022.06460.PTDC).  A.L.W. acknowledges support from ERC Consolidator Grant No.~865768 AEONS.  Computations were carried out on the HELIOS cluster on dedicated
nodes funded via this grant.

\section*{Data Availability}

 The posterior samples and scripts to make the plots in this paper are available in a Zenodo repository \citep{ChunHuang2024}



\bibliographystyle{mnras}
\bibliography{example} 

\bsp	
\label{lastpage}
\end{document}